\documentclass[twocolumn]{aastex631}

\usepackage{verbatim}
\usepackage[T1]{fontenc} % for Icelandic Characters
\usepackage[utf8]{inputenc} % for Icelandic Characters 
\shorttitle{AASTeX v6.3.1 Sample article}
\shortauthors{Powers et al.}
\graphicspath{{./}{figures/}}
\begin{document}

\title{TOI-3785~b: A Low-Density Neptune Orbiting an M2-Dwarf Star}

\author[0000-0002-5300-5353]{Luke C. Powers}
\affil{Department of Astronomy \& Astrophysics, 525 Davey Laboratory, The Pennsylvania State University, University Park, PA, 16802, USA}
\affil{Center for Exoplanets and Habitable Worlds, 525 Davey Laboratory, The Pennsylvania State University, University Park, PA, 16802, USA}

\author[0000-0002-2990-7613]{Jessica Libby-Roberts}
\affil{Department of Astronomy \& Astrophysics, 525 Davey Laboratory, The Pennsylvania State University, University Park, PA, 16802, USA}
\affil{Center for Exoplanets and Habitable Worlds, 525 Davey Laboratory, The Pennsylvania State University, University Park, PA, 16802, USA}

\author[0000-0002-9082-6337]{Andrea S.J.\ Lin}
\affil{Department of Astronomy \& Astrophysics, 525 Davey Laboratory, The Pennsylvania State University, University Park, PA, 16802, USA}
\affil{Center for Exoplanets and Habitable Worlds, 525 Davey Laboratory, The Pennsylvania State University, University Park, PA, 16802, USA}

\author[0000-0003-4835-0619]{Caleb I. Ca\~nas}
\altaffiliation{NASA Postdoctoral Program Fellow}
\affil{NASA Goddard Space Flight Center, 8800 Greenbelt Road, Greenbelt, MD 20771, USA}
\affil{Department of Astronomy \& Astrophysics, 525 Davey Laboratory, The Pennsylvania State University, University Park, PA, 16802, USA}
\affil{Center for Exoplanets and Habitable Worlds, 525 Davey Laboratory, The Pennsylvania State University, University Park, PA, 16802, USA}

\author[0000-0001-8401-4300]{Shubham Kanodia}
\affil{Earth and Planets Laboratory, Carnegie Institution for Science, 5241 Broad Branch Road, NW, Washington, DC 20015, USA}
\affil{Department of Astronomy \& Astrophysics, 525 Davey Laboratory, The Pennsylvania State University, University Park, PA, 16802, USA}
\affil{Center for Exoplanets and Habitable Worlds, 525 Davey Laboratory, The Pennsylvania State University, University Park, PA, 16802, USA}

\author[0000-0001-9596-7983]{Suvrath Mahadevan}
\affil{Department of Astronomy \& Astrophysics, 525 Davey Laboratory, The Pennsylvania State University, University Park, PA, 16802, USA}
\affil{Center for Exoplanets and Habitable Worlds, 525 Davey Laboratory, The Pennsylvania State University, University Park, PA, 16802, USA}
\affil{ETH Zurich, Institute for Particle Physics \& Astrophysics, Switzerland}

\author[0000-0001-8720-5612]{Joe P.\ Ninan}
\affil{Department of Astronomy and Astrophysics, Tata Institute of Fundamental Research, Homi Bhabha Road, Colaba, Mumbai 400005, India}

\author[0000-0001-7409-5688]{Guðmundur Stefánsson} 
\affil{NASA Sagan Fellow}
\affil{Department of Astrophysical Sciences, Princeton University, 4 Ivy Lane, Princeton, NJ 08540, USA}

\author[0000-0002-5463-9980]{Arvind F.\ Gupta}
\affil{Department of Astronomy \& Astrophysics, 525 Davey Laboratory, The Pennsylvania State University, University Park, PA, 16802, USA}
\affil{Center for Exoplanets and Habitable Worlds, 525 Davey Laboratory, The Pennsylvania State University, University Park, PA, 16802, USA}

%Mark

\author[0000-0002-7227-2334]{Sinclaire Jones}
\affil{Department of Astronomy, The Ohio State University, 4055 McPherson Laboratory, Columbus, OH 43210, USA}
\affil{Department of Astrophysical Sciences, Princeton University, 4 Ivy Lane, Princeton, NJ 08540, USA}

\author[0000-0002-4475-4176]{Henry A. Kobulnicky}
\affil{Department of Physics \& Astronomy, University of Wyoming, Laramie, WY 82070, USA}

\author[0000-0002-0048-2586]{Andrew Monson}
\affil{Steward Observatory, The University of Arizona, 933 N.\ Cherry Avenue, Tucson, AZ 85721, USA}

\author[0000-0001-9307-8170]{Brock A. Parker}
\affil{Department of Physics \& Astronomy, University of Wyoming, Laramie, WY 82070, USA}

\author[0000-0002-5817-202X]{Tera N. Swaby}
\affil{Department of Physics \& Astronomy, University of Wyoming, Laramie, WY 82070, USA}

\author[0000-0003-4384-7220]{Chad F. Bender}
\affil{Steward Observatory, The University of Arizona, 933 N.\ Cherry Avenue, Tucson, AZ 85721, USA}

\author[0000-0001-9662-3496]{William D. Cochran}
\affil{McDonald Observatory and Center for Planetary Systems Habitability
The University of Texas, Austin Texas USA}
%List NSF grant AST 2108801

%Sam

\author[0000-0003-1263-8637]{Leslie Hebb}
\affil{Physics Department, Hobart and William Smith Colleges, 300 Pulteney Street, Geneva, NY 14456, USA}
\affil{Department of Astronomy, Cornell University, 245 East Ave, Ithaca, NY 14850, USA}

\author[0000-0001-5000-1018]{Andrew J. Metcalf}
\affil{Air Force research laboratory, Kirtland AFB, Albuquerque NM 87117}

\author[0000-0003-0149-9678]{Paul Robertson}
\affil{Department of Physics \& Astronomy, University of California Irvine, Irvine, CA 92697, USA}

\author[0000-0002-4046-987X]{Christian Schwab}
\affil{School of Mathematical and Physical Sciences, Macquarie University, Balaclava Road, North Ryde, NSW 2109, Australia}
%Ryan

\author[0000-0001-9209-1808]{John Wisniewski}
\affil{George Mason University  Department of Physics \& Astronomy, 4400 University Drive, MS 3F3, Fairfax, VA 22030, USA.}

\author[0000-0001-6160-5888]{Jason T.\ Wright}
\affil{Department of Astronomy \& Astrophysics, 525 Davey Laboratory, The Pennsylvania State University, University Park, PA, 16802, USA}
\affil{Center for Exoplanets and Habitable Worlds, 525 Davey Laboratory, The Pennsylvania State University, University Park, PA, 16802, USA}
\affil{Penn State Extraterrestrial Intelligence Center, 525 Davey Laboratory, The Pennsylvania State University, University Park, PA, 16802, USA}



\begin{abstract}
Using both ground-based transit photometry and high-precision radial velocity (RV) spectroscopy, we confirm the planetary nature of TOI-3785~b. This transiting Neptune orbits an M2-Dwarf star with a period of $\sim$4.67 days, a planetary radius of $5.14 \pm 0.16$ $ R_\oplus$, a mass of 14.95$^{+4.10}_{-3.92}$ $M_\oplus$, and a density of $\rho =0.61^{+0.18}_{-0.17}$ $g/cm^3$. TOI-3785~b belongs to a rare population of Neptunes $(4~R_\oplus < R_p < 7~R_\oplus)$ orbiting cooler, smaller M-dwarf host stars, of which only $\sim$ 10 have been confirmed. By increasing the number of confirmed planets, TOI-3785~b offers an opportunity to compare similar planets across varying planetary and stellar parameter spaces. Moreover, with a high transmission spectroscopy metric (TSM) of $\sim$150 combined with a relatively cool equilibrium temperature of $T_{\rm{eq}} = 582 \pm 16$ K and an inactive host star, TOI-3785~b is one of the more promising low-density M-dwarf Neptune targets for atmospheric follow-up. Future investigation into atmospheric mass loss rates of TOI-3785~b may yield new insights into the atmospheric evolution of these low-mass gas planets around M-dwarfs.
\end{abstract}

%\keywords{planets and satellites: detection, composition; planetary systems; stars: fundamental parameters; methods: statistical;}

\section{Introduction} \label{sec:intro}

The success of the \emph{Kepler} \citep{Kepler} and \textit{TESS} missions \citep{RickerTESS} have produced a catalog of over 5000 confirmed exoplanets. Multiple studies have leveraged these detections to derive planetary occurrence rates across a wide range of parameter spaces. Planetary occurrence rates around M-dwarf stars (the most common spectral type in our galaxy) are of particular interest. Using \emph{Kepler}, \citet{dressing.mdwarf.occurrence} found that small, short-period planets (such as super-Earths and sub-Neptunes, $1.4~R_\oplus<R_p<4~R_\oplus)$ are more common around M-dwarfs than that of the Neptune- and Jupiter-sized planets. According to the NASA Exoplanet Archive \citep{Archive} there are only $\sim$10 transiting planets within the Neptune radii bounds $(4~R_\oplus < R_p < 7~R_\oplus)$ with confirmed masses orbiting M-dwarf stars, significantly less than the terrestrial population. The processes by which these larger Neptunes orbiting low-mass stars form is still an open question---one that requires a larger sample of planets to answer. Discovering and characterizing more of these planets with precise radius and mass measurements will continue to aid efforts to quantify occurrence and understand the specific mechanisms behind M-dwarf planetary formation. 

We present a new planet inhabiting this sparsely-populated M-dwarf Neptune parameter space, TOI-3785~b. We used a combination of ground-based photometric (transit) and spectroscopic (radial velocity) follow-up to confirm this \textit{TESS} discovered planet which we describe in Section~\ref{sec:observations}. Using stellar spectra, we update the stellar parameters (Section~\ref{sec:stellarparams}) confirming that TOI-3785 is an inactive M-dwarf. We derive precise mass and radius measurements for TOI-3785~b in Section~\ref{sec:analysis}. In Section~\ref{sec:discussion} we highlight TOI-3785~b's place across a variety of stellar and planetary parameters and discuss its potential for various in-depth studies into comparative planetology. We conclude and summarize this work in Section~\ref{sec:summary}. %Additionally, we update the stellar parameters of the M-dwarf host star and verify its inactive nature. We find that TOI-3785~b is a slightly larger and more massive clone of GJ 3470~b \citep{Stefansson2022_GJ3470b}, orbiting a nearly identical M-dwarf star at a very similar distance. This makes TOI-3785~b an intriguing target for atmospheric comparison with the well-studied GJ 3470~b. %Given the relatively bright magnitude (J = 11.05 $\pm$ 0.03, see \autoref{tab:SP} for list of magnitudes), TOI-3785~b possesses one of the highest transmission spectroscopy metrics \citep[TSM;][]{kempton.tsm} for planets with temperatures cooler than 600 K.  

%Formation and evolutionary timelines for this target and other similar Neptunes are also considered. Estimations of core mass and atmospheric composition reveal possible routes TOI-3785~b could have taken to become the short period Neptune we see today.  

%In this paper, we report our photometric and radial velocity observations of TOI-3785~b in Section~\ref{sec:observations}. The updated stellar parameters derived from this fitted data are listed in Section~\ref{sec:stellarparams}. Section~\ref{sec:analysis} summarizes our analysis of the photometric and spectroscopic data, confirming the planetary nature of TOI-3785~b. Section~\ref{sec:discussion} discusses TOI-3785~b's place among previously confirmed M-dwarf planetary systems including a discussion on atmospheric characterization. We summarize this work in Section~\ref{sec:summary}.  

\section{Observations} \label{sec:observations}
\subsection{TESS Photometry}

\begin{figure*}[!h]
    % transits
    \begin{minipage}[b]{0.5\linewidth}
    \centering
    \includegraphics[width=\textwidth]{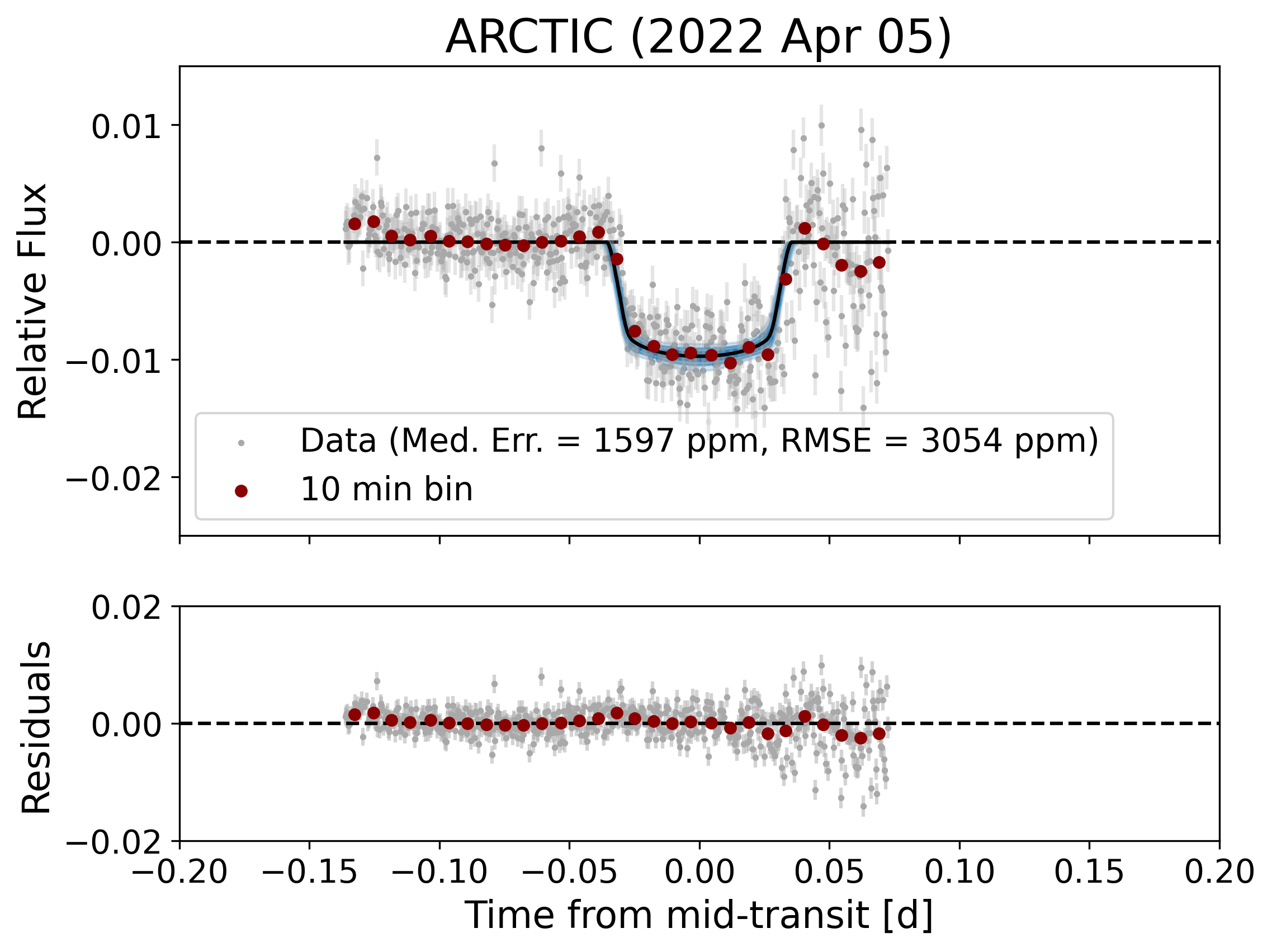}
    \end{minipage}
    \begin{minipage}[b]{0.5\linewidth}
    \centering
    \includegraphics[width=\textwidth]{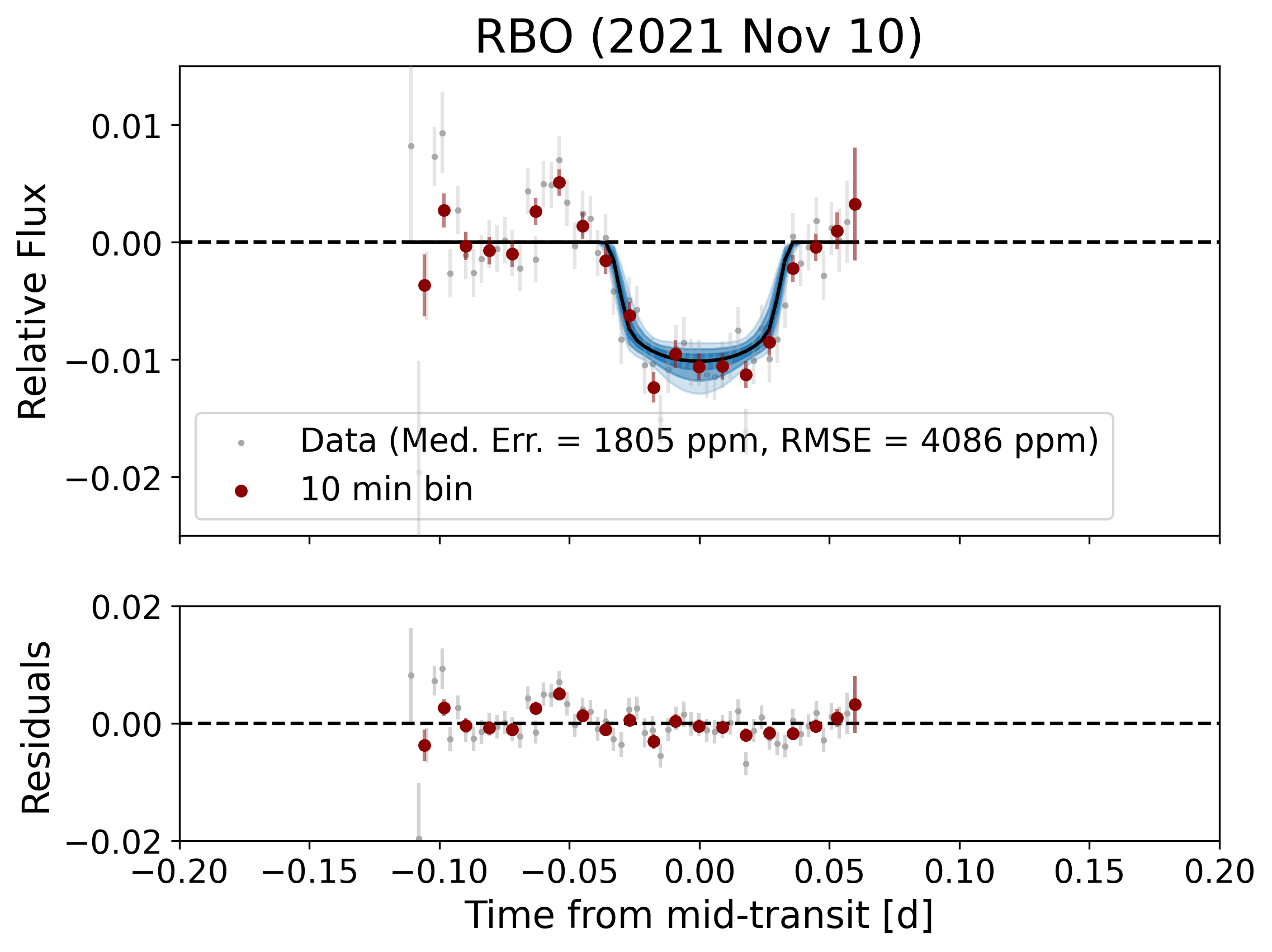}
    \end{minipage}
    % transits 2
    \begin{minipage}[b]{0.5\linewidth}
    \centering
    \includegraphics[width=\textwidth]{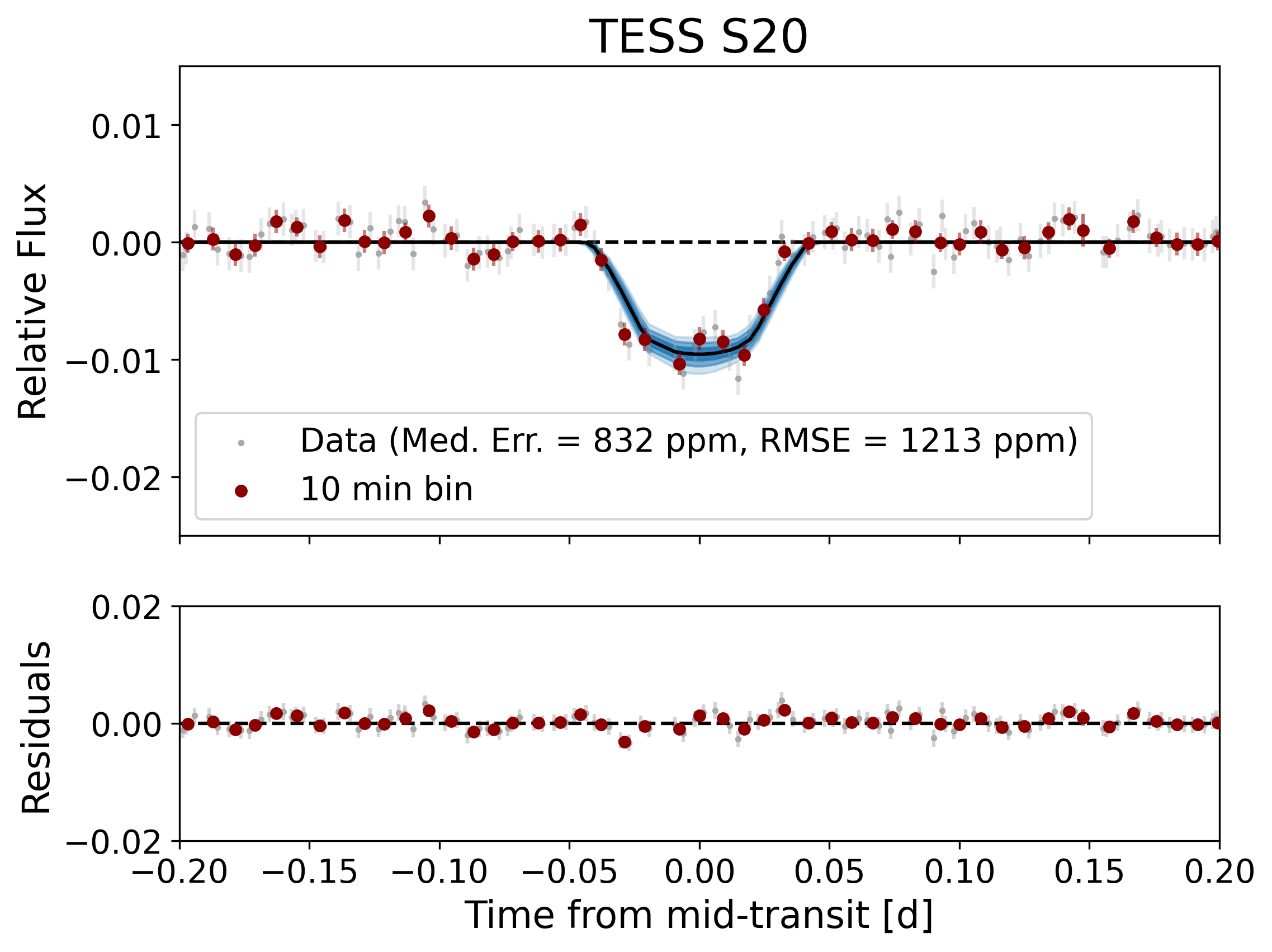}
    \end{minipage}
    \begin{minipage}[b]{0.5\linewidth}
    \centering
    \includegraphics[width=\textwidth]{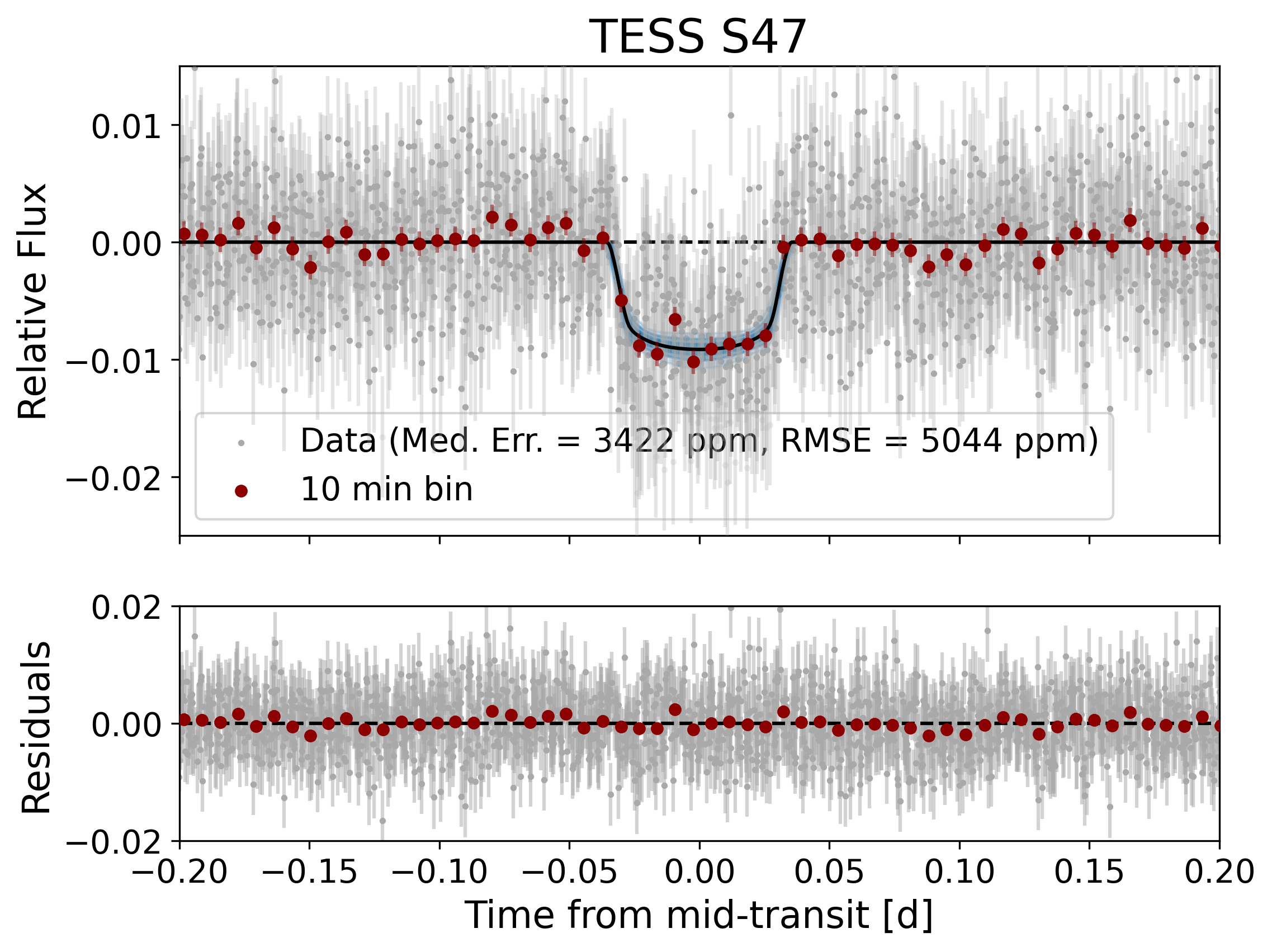}
    \end{minipage}
\caption{The transit observations conducted for TOI-3785~b. In each case, we fit a model to the lightcurve via \texttt{exoplanet} (Section~\ref{sec:analysis}). The best-fit model is shown in black, with the 1, 2, and $3\sigma$ confidence intervals in progressively lighter shades of blue. A 10-minute bin of our data is included as red points in each panel. It is also important to note that the reported median errors are dependent on their respective exposure time of the instrument used. The subpanels show the ARCTIC transit observed in the Semrock filter (top left), the RBO transit observed in Bessell I (top right), TESS Sector 20 data taken at long cadence (1800 seconds) (bottom left), and TESS Sector 47 data taken at short cadence (120 seconds) (bottom right). }
\label{fig:LC}
\end{figure*}

TOI-3785 (\autoref{tab:SP}) was first observed in \textit{TESS} Sector 20, from 2019 December 24 to 2020 January 20. Similar to the TOI-1899 \citep{Canas_2020} and TOI-3629 \citep{Canas_2022} systems, we identified TOI-3785~b as a planetary candidate using a custom pipeline to search for transiting candidates in short and long-cadence TESS data. This target was independently identified by the Quick Look Pipeline \citep[QLP;][]{QLP} when a 4.67 day transiting signal was flagged during an observation in long cadence mode (1800-second exposure). An identical periodic signal from TOI-3785 was again observed by \textit{TESS} in Sector 47 from 2021 December 30 to 2022 January 28 with a two-minute exposure time. We retrieved both long and short cadence sectors using the \texttt{lightkurve} package \citep{Lightkurve}. The Pre-search Data Conditioning Simple Aperture Photometry \citep[PDCSAP;][]{Jenkins2016_tess_spoc} flux was used during our analysis \citep{2020RNAAS...4..201C}. We show this photometry along with the best-fit model from our joint fit in \autoref{fig:LC}.

\begin{deluxetable*}{llrr}[htbp]
\centering
\caption{Stellar Parameters}
\label{tab:SP}
\tabletypesize{\footnotesize} % \scriptsize
\tablehead{\colhead{Parameter} & \colhead{Description} & \colhead{Value} & \colhead{Source}}
\startdata
\multicolumn{2}{l}{\textbf{Main Identifiers:}} \\
\quad TOI &TESS Object of Interest &3785 & ExoFOP-TESS \citet{NEX}\\
\quad TIC &TESS Input Catalogue &458419328 & ExoFOP-TESS \citet{NEX}\\
\quad 2MASS &... &J08433613+6304413 & ExoFOP-TESS \citet{NEX}\\
\quad Gaia DR3 &... &1044013542142711296 & ExoFOP-TESS \citet{NEX}\\
\quad APASS &... &59229225 &ExoFOP-TESS \citet{NEX} \\
\multicolumn{2}{l}{\textbf{Equatorial Coordinates:}}\\
\quad $\alpha_{J2000}$&Right Ascension (RA) &8:43:36 & ExoFOP-TESS \citet{NEX}\\
\quad $\delta_{J2000}$&Declination (DEC) &+65:03:41 & ExoFOP-TESS \citet{NEX}\\
\multicolumn{2}{l}{\textbf{Proper Motion:}}\\
\quad $\mu_\alpha$&Proper motion (RA) &-42.86 $\pm$ 0.01 & GAIA (DR3) \citep{GDR3} \\
\quad $\mu_\delta$&Proper motion (DEC) &-16.95 $\pm$ 0.01 & GAIA (DR3) \citep{GDR3}\\
\multicolumn{2}{l}{\textbf{Distance and maximum extinction:}}\\
\quad $d$ &  Geometric distance (pc)  & $79.4\pm0.1$ & \citet{Bailer-Jones2021}\\
\quad \(A_{V,max}\) & Maximum visual extinction & $0.03$ & \citet{Green2019}\\
\multicolumn{2}{l}{\textbf{Magnitudes:}}\\
\quad TESS &TESS mag &12.496 $\pm$0.007 & ExoFOP-TESS \citet{NEX}\\
% u &Sloan u mag & 17.92 \pm 0.01 & SDSS\\
% g &Sloan g mag &15.475 \pm 0.004 & SDSS\\
% r &Sloan r mag &14.074 \pm 0.003 & SDSS\\
%\quad i &Sloan \textit{i} mag &13.223 $\pm$ 0.001 & SDSS\\
%\quad z &Sloan \textit{z} mag &12.406 $\pm$ 0.004 & SDSS\\
\quad g &PS1 \textit{g'} mag &15.244 $\pm$ 0.013 & PS1 \cite{Chambers2016,Magnier2020}\\
\quad r &PS1 \textit{r'} mag &14.076 $\pm$ 0.008 & PS1 \cite{Chambers2016,Magnier2020}\\
\quad y & PS1 \textit{y'} mag &12.248 $\pm$ 0.022 & PS1 \cite{Chambers2016,Magnier2020}\\
\quad J &2M J mag &11.051 $\pm$ 0.026 & 2MASS \citep{cutri_2mass_2033}\\
\quad H &2M H mag &10.387 $\pm$ 0.029 & 2MASS \citep{cutri_2mass_2033}\\
\quad K &2M K mag &10.165 $\pm$ 0.022 & 2MASS \citep{cutri_2mass_2033}\\
\quad W1 & WISE1 mag & $10.034 \pm 0.023$ & WISE \citep{Wright2010}\\
\quad W2 &  WISE2 mag & $9.966 \pm 0.019$ & WISE \citep{Wright2010}\\
\quad W3 &  WISE3 mag & $9.860 \pm 0.045$ & WISE \citep{Wright2010}\\
\hline %\bottomrule
\enddata
\end{deluxetable*}

\subsection{Ground-based photometric follow-up}

\subsubsection{Red Buttes Observatory 0.6 m} %(airmass 2-3) 
One transit of TOI-3785~b was observed on 2021 November 11 using the 0.6 m telescope at the Red Buttes Observatory (RBO) in Wyoming \citep{Kasper}. We observed TOI-3785~b using the Bessell I filter at an exposure time of 240 seconds, from an airmass of 1.26 to 1.08. The post-transit observations were cut short due to increased cloud cover (\autoref{fig:LC}). 

\subsubsection{ARC 3.5 m Telescope}
We obtained one transit of TOI-3785~b on the night of 2022 April 5 using the Astrophysical Research Consortium (ARC) Telescope Imaging Camera \citep[ARCTIC;][]{Huehnerhoff2016} at the ARC 3.5 m Telescope at Apache Point Observatory (APO). This target was observed with ARCTIC's narrow-band Semrock filter \citep[between 842 and 873 nm;][]{Stefansson2017_diffuser, Stefansson2018_diffuserSPIE}, with an exposure time of 56 seconds, in the quad-amplifier, fast readout mode, and with 4 $\times$ 4 on-chip binning mode in effect. Relatively photometric skies and the use of this narrow-band Semrock filter (designed to avoid regions of telluric water absorption), enabled us to obtain high-precision photometry even at a significant airmass change (airmass 1.38 to 3.81) over the entirety of the transit event. The increasing airmass towards the end (airmass $>$ 3) resulted in significant scatter in the post-transit baseline (\autoref{fig:LC}).

\subsubsection{NESSI at WIYN} 
The NN-EXPLORE Exoplanet Stellar Speckle Imager \citep[NESSI;][]{NESSI} is mounted on the WIYN 3.5 m telescope at Kitt Peak National Observatory (KPNO). We used NESSI speckle photometry to search for faint stellar targets in close proximity to TOI-3785 that may contaminate the primary, or introduce additional photometric error. We observed this target on 2022 April 17 in the Sloan z$^{\prime}$ filter (865 nm - 960 nm). 

\begin{figure}[!htbp]
\begin{center}
    %\textbf{}\par\medski
  \includegraphics[width=\columnwidth]{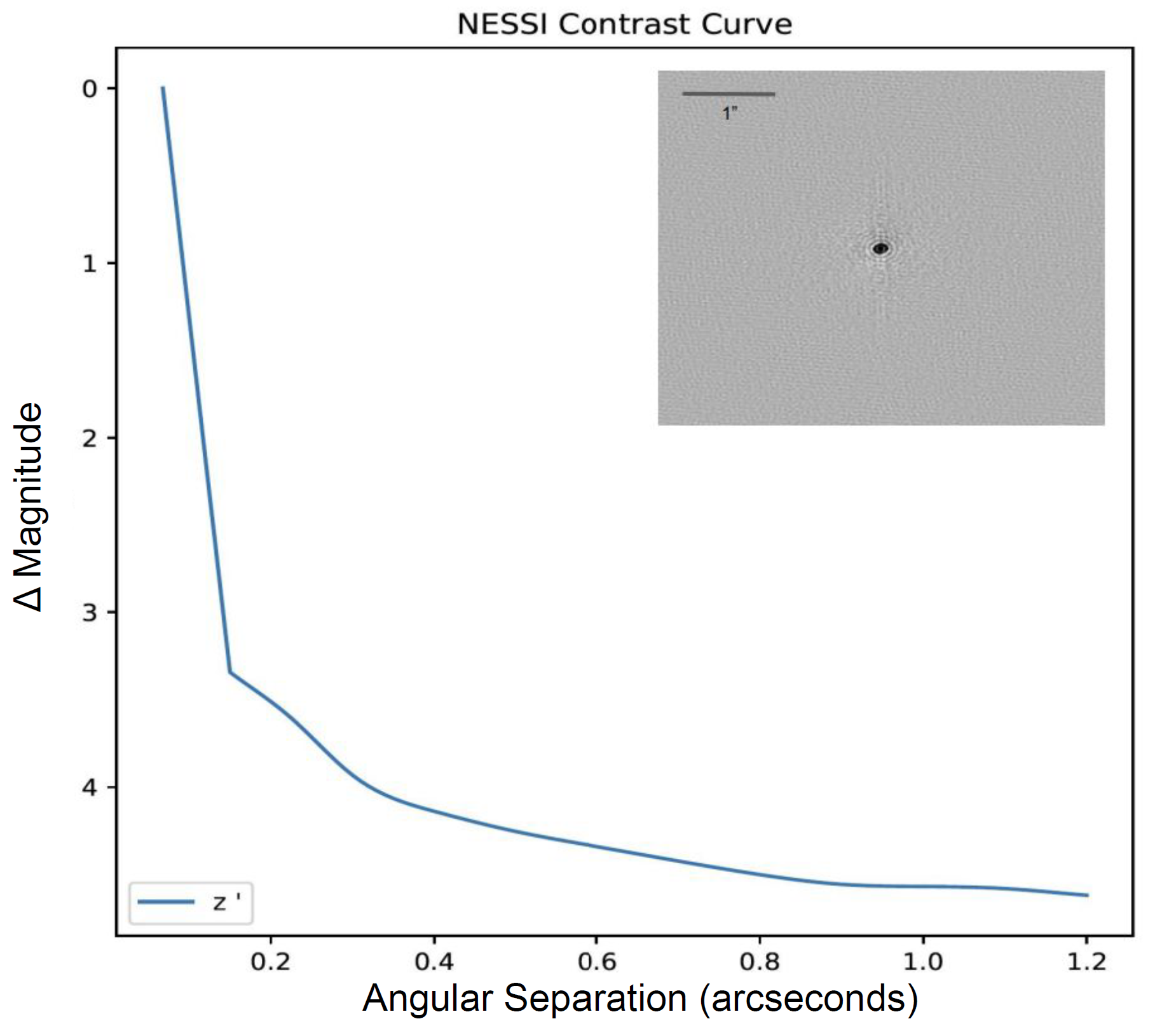}
  \end{center}
  \caption{5$\sigma$ contrast curve from NESSI in Sloan z', showing no evidence of faint companions within 1.2 arcsecond. Inset: The 2D reconstructed image of the 1.2-arcsecond region surrounding TOI-3785.}
  \label{fig:NESSI}
\end{figure}

The 5-sigma contrast curve in \autoref{fig:NESSI} reveals no bright ($\Delta$z$^{\prime}$ mag $<$ 4) stellar companions within a 0.3 - 1.2 arcsecond range of TOI-3785. We also include the 2D NESSI speckle image for TOI-3785 as an inset.

Additionally, we use \emph{Gaia} Data Release 3 \citep[DR3;][]{GDR3} to further rule-out stellar companions within a 25\arcsec~range. According to \citet{2018AJ....156..259Z}, Gaia has the capabilities to recover 93\% of targets at a distance $>$2\arcsec. In TOI-3785's case, Gaia DR3 reveals the closest object at 26\arcsec. Therefore, considering data from NESSI and Gaia, we can conclude that no source of significant photometric dilution is present from nearby stellar companions.

\subsection{Radial Velocity follow-up}

\subsubsection{The Habitable-zone Planet Finder}
The Habitable-zone Planet Finder \citep[HPF;][]{MHPF16,MHPF14} is designed to obtain high-precision near-infrared (808 nm -- 1278 nm) radial velocity observations. Located on the 10 m Hobby-Eberly Telescope \citep[HET;][]{Ramsey1998_HET, Hill2021_HET} in Texas, this spectrograph is rigorously environmentally-controlled \citep{Stefansson2016_hpf_environment} and fiber-fed, allowing for simultaneous science and sky observations \citep{KanodiaFIBERS}. From the raw HPF data, we correct for bias noise, cosmic rays, and non-linearity using \texttt{HxRGproc} \citep{ninan}.

We apply a modified version of the SpEctrum Radial Velocity AnaLyser pipeline \citep[\texttt{SERVAL;}][]{SERVAL}, as outlined in \citet{Metcalf}, to derive the binned radial velocity (RV) points. To accomplish this, \texttt{SERVAL} combines all observations of TOI-3785 to extract a master spectrum \citep{2012ApJS..200...15A} after first identifying and masking telluric and sky emission lines. \texttt{SERVAL} then fits this template to each individual spectrum by shifting it in wavelength space to minimize $\chi^{2}$. We use the python package \texttt{barycorrpy} \citep{2018RNAAS...2....4K} to further correct for barycentric motion.

\begin{figure*}[!ht]
\begin{center}
\hspace*{-1cm}
    % \textbf{EXOPLANET RV RUN (HPF and NEID)}\par\medskip
  \includegraphics[scale=0.8]{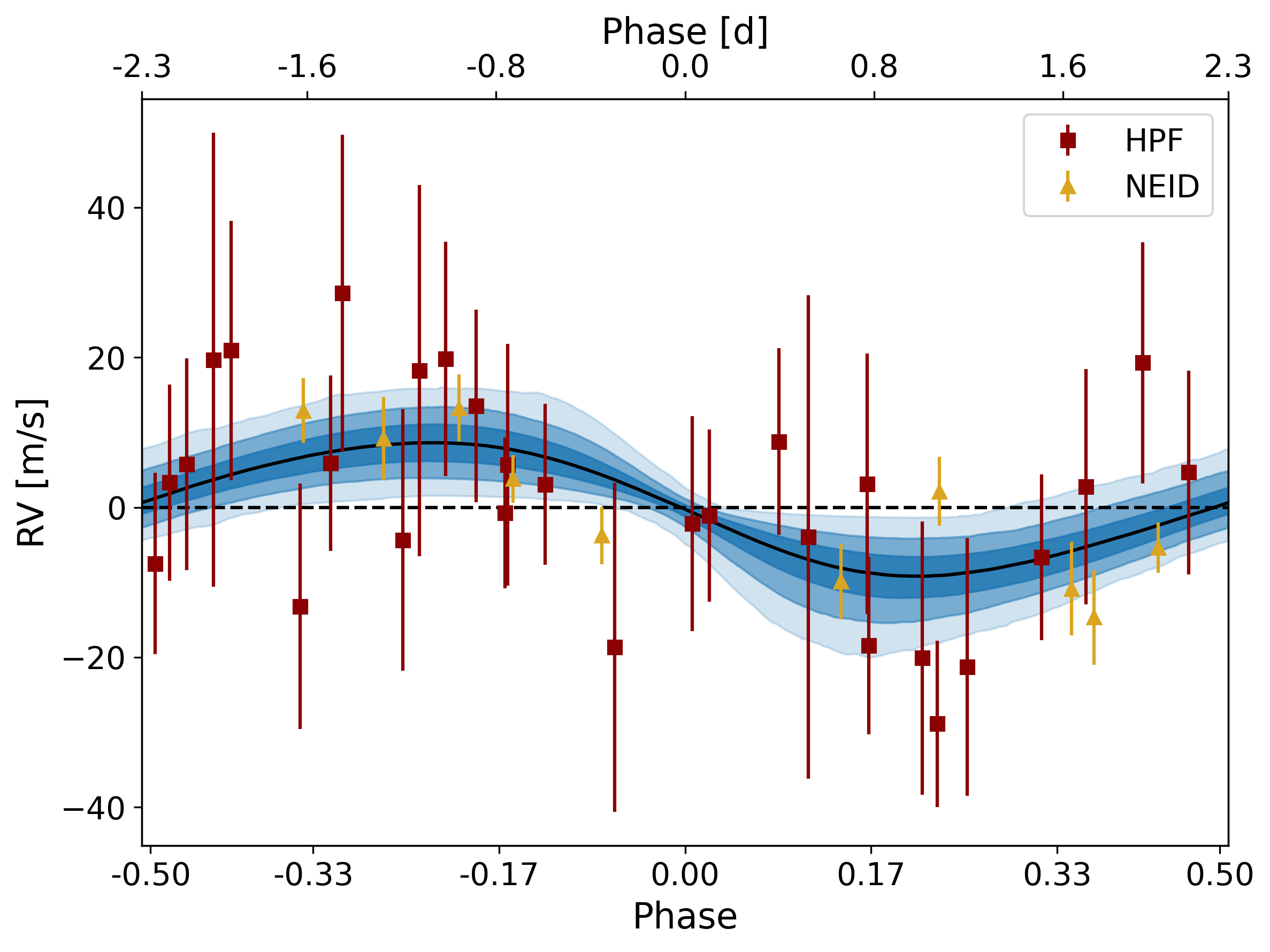}
  \end{center}
  \caption{Phase folded RV points from both HPF (dark red) and NEID (gold) with the best-fit model plotted in blue along with the 1, 2, and 3 $\sigma$ contours (similar to \autoref{fig:LC}) generated from the \texttt{exoplanet} joint fit. }
\label{fig:RVcurve}
\end{figure*}

We observed TOI-3785 with HPF for 34 visits, with most visits consisting of two 15-minute exposures per night that were then binned, between 2020 November 4 and 2022 April 19. A median Signal-to-Noise ratio (S/N) of 69 was calculated at a wavelength of 1070 nm. Of the 34 collected RV points, 29 were kept for the final analysis. Discarded points were done so on the grounds of either unideal weather conditions or significant deviation from the average S/N. Binned RV points along with their errors are listed under \autoref{tab:HPFRVS}, and the final binned HPF RVs are plotted as dark red points in \autoref{fig:RVcurve}.

\begin{deluxetable}{lrrr}[htbp]
\centering
\caption{HPF RV Points of the high S/N points used in the analysis}
\label{tab:HPFRVS}
\tabletypesize{\footnotesize}
\tablehead{\colhead{BJD$_{\rm{TDB}}$} & \colhead{RV (m/s)} & \colhead{$\sigma$ (m/s)} & \colhead{S/N}}
\startdata
2459157.97334\tablenotemark{a} & 23.98    & 16.08           & 80  \\
2459182.02270\tablenotemark{a} & 24.38    & 30.28           & 45  \\
2459320.64060\tablenotemark{a} & -15.41   & 18.21           & 72  \\
2459513.00857\tablenotemark{a} & 7.44     & 15.69           & 83  \\
2459514.00493                  & 25.61    & 17.30           & 56  \\
2459515.98987                  & 2.52     & 14.34           & 65  \\
2459547.90755                  & 3.92     & 10.06           & 91  \\
2459557.00253                  & 24.52    & 15.63           & 61  \\
2459564.87333                  & 9.37     & 13.59           & 68  \\
2459565.85709                  & 10.60    & 11.69           & 80  \\
2459570.84041                  & 0.34     & 17.46           & 52  \\
2459572.83906                  & 7.83     & 17.39           & 56  \\
2459575.96782                  & 10.38    & 16.12           & 59  \\
2459577.94491                  & -16.59   & 17.21           & 55  \\
2459593.77216                  & -8.50    & 16.38           & 58  \\
2459604.88419                  & 3.60     & 11.49           & 82  \\
2459626.67754                  & 33.26    & 21.17           & 46  \\
2459628.68188\tablenotemark{a} & 0.75     & 32.24           & 46  \\
2459629.68568                  & -1.94    & 11.05           & 85  \\
2459630.68258                  & 10.46    & 14.14           & 67  \\
2459631.68481\tablenotemark{a} & 22.92    & 24.79           & 57  \\
2459644.63493                  & 8.00     & 13.10           & 72  \\
2459647.64145                  & -13.73   & 11.86           & 82  \\
2459678.67461                  & 18.21    & 12.85           & 75  \\
2459680.66011                  & -24.19   & 11.09           & 86  \\
2459683.64781                  & 7.76     & 10.76           & 88  \\
2459684.65200                  & 13.47    & 12.45           & 76  \\
2459686.64437                  & -2.79    & 12.09           & 79  \\
2459688.62032                  & -13.99   & 21.92           & 45  \\
\enddata
\tablenotetext{a}{Single 15-minute exposure during the nightly visit.}
\end{deluxetable}

\subsubsection{NEID}
NEID \citep{SchwabNEID, Halverson2016_errorbudget} is a high-resolution (R $\sim$ 110,000) spectrograph located on the WIYN 3.5 m telescope at KPNO. NEID covers optical/near-infrared wavelengths ranging from 380 nm to 930 nm. We observed TOI-3785~between 2021 November 10 and 2022 May 16, obtaining 10 RV points with NEID in high-resolution (HR) mode. As NEID allows for longer exposure times than HPF, we obtained a single spectrum per visit with an exposure time of 1800 seconds, resulting in a median S/N of 15 at 850 nm. The raw spectra were reduced through the NEID Data Reduction Pipeline\footnote{\url{https://neid.ipac.caltech.edu/docs/NEID-DRP/}}, and we retrieved the Level-2 2D extracted 
spectra.\footnote{\url{https://neid.ipac.caltech.edu/search.php}} We derived RVs using a modified \textit{SERVAL} pipeline designed specifically for NEID data \citep{Gummi3470}. NEID RVs and errors are reported in \autoref{tab:NEIDRVs} and plotted in gold in \autoref{fig:RVcurve}.

\begin{deluxetable}{lrrr}[htbp]
\centering
\caption{NEID RV Points}
\label{tab:NEIDRVs}
\tabletypesize{\footnotesize}
\tablehead{\colhead{BJD$_{\rm{TDB}}$} & \colhead{RV (m/s)} & \colhead{$\sigma$ (m/s)} & \colhead{S/N}}
\startdata
2459529.01052 & 15.7     & 5.26           & 15  \\
2459532.01789 & -4.52    & 3.91           & 20  \\
2459533.01626 & 11.74    & 5.07           & 16  \\
2459533.91777 & 12.63    & 3.69           & 21  \\
2459538.97354 & -9.06    & 4.43           & 18  \\
2459569.04250 & -9.32    & 7.27           & 11  \\
2459586.75088 & -8.23    & 5.89           & 14  \\
2459619.89833 & 7.11     & 5.43           & 15  \\
2459629.91089 & 0.82     & 7.47           & 11  \\
2459715.67442 & 18.02    & 6.51           & 13  \\
\enddata
\end{deluxetable}
\section{Stellar Parameters} \label{sec:stellarparams}
We used the HPF spectra and \texttt{HPF-SpecMatch} \citep[][]{2020AJ....159..100S} to derive the effective temperature ($T_{\rm{eff}}$), metallicity (Fe/H), $v \sin{i}$, and $\log{g}$ priors for the host star, TOI-3785. Based on \citet{2017ApJ...836...77Y}, \texttt{HPF-SpecMatch} uses a spectral database of well-characterized stellar targets with high S/N HPF observations comparing each star to that of TOI-3785. By creating a composite of library spectra and minimizing the $\chi^{2}$ of the composite, we obtain best-fit values for each parameter. Uncertainties in the spectroscopic parameters were then determined from cross-validation estimates (for additional details see \citet{2020AJ....159..100S}). We estimate the following stellar priors, $T_{eff} = 3576\pm88$ K, $log(g) = 4.747\pm0.0458$, and $Fe/H = 0.099\pm0.117$ (\autoref{tab:Parameters}).

\begin{deluxetable*}{lrr}[htbp]
\centering
\caption{TOI-3785~b System Parameters}
\label{tab:Parameters}
\tabletypesize{\footnotesize} % \scriptsize
\tablehead{\colhead{Parameter} & \colhead{Label (Units)} & \colhead{Value}}
\startdata
\textbf{Orbital Parameters:} \\
\quad RV Semi-Amplitude &$K$ (m/s) &9.24 $\pm$ 2.68 \\
\quad Orbital Period &$P$ (days) &$4.6747373\pm{0.0000038}$ \\
\quad Transit Midpoint &$T_0$ (BJD) &2458861.49553$^{+0.00060}_{-0.00058}$ \\
\quad Scaled Radius &$R_p/R_*$ &0.0962$\pm$ 0.0017 \\
\quad Scaled Semi-major Axis &$a/R_*$ &18.89$^{+0.45}_{-0.44}$ \\
\quad Impact Parameter &$b$ &0.60$^{+0.02}_{-0.03}$ \\
\hline %\midrule
\textbf{Planetary Parameters: } \\
\quad Eccentricity &$e$ &$ 0.11^{+0.10}_{-0.08}$ $(2\sigma<0.26)$ \\
\quad Inclination &$i$ (degrees) &88.1$\pm 0.01$ \\
\quad Omega &$\omega$ (degrees) &96.26$^{+51.25}_{-143.93}$ \\
\quad Transit Duration &$T_{\rm{Dur}}$ (days) &0.071$\pm$ 0.001 \\
\quad Transit Depth & $(R_p/R_*)^2$ (ppm) &$9254 \pm 3$ \\
\quad Mass &$M_p$ ($M_\oplus$) &14.95$^{+4.10}_{-3.92}$ \\
\quad Radius &$R_p$ ($R_\oplus$) &5.14 $\pm 0.16$ \\
\quad Density &$\rho_p$ (g/cm$^{3}$) &0.61$^{+0.18}_{-0.17}$ \\
\quad Semi-Major Axis &$a$ (AU) &0.043$\pm 0.001$ \\
\quad Isolation &$S_p$ ($S_\oplus$) &19.1$\pm 2.0$ \\
\quad Equilibrium Temperature &$T_{eq}$ (K) &582$\pm 16$ \\
\hline %\bottomrule
\textbf{Stellar Parameters:} \\
\quad Mass &$M_*$ ($M_\odot$)&$0.52\pm{0.02}$\\
\quad Radius &$R_*$ ($R_\odot$)&$0.50\pm0.01$\\
%\quad Effective Temperature &$T_{eff}$ (Kelvin) &3575$^{+86}_{-85}$ & & & \\
%\quad Luminosity & $L_\odot$(Solar Luminosity) &0.036$^{+0.0037}_{-0.0037}$ & & & \\
%\hline %\bottomrule
%\textbf{Stellar Parameters: [SED]  } & & & & & \\
%\quad Mass &$M_\odot$(Solar Mass)&0.48$\pm{0.03}$\\
%\quad Radius &$R_\odot$(Solar Radii)&0.45$\pm{0.02}$\\
\quad Luminosity & $L_*$ ($L_\odot$) &$0.0367^{+0.0008}_{-0.0009}$ \\
%\textbf{Stellar Parameters: [Spec-match]  } & & & & & \\
\quad Effective Temperature &$T_{\rm{eff}}$ (Kelvin) &$3576\pm88$ K \\
\quad Surface gravity &  $\log{g}$ (cgs) & 4.747$\pm0.0458$ \\
\quad Rotational velocity & $v \sin{i}$ (km/s)  &$<2$ \\
\quad Metallicity &  [Fe/H] (dex) &$0.099\pm0.117$ \\
\quad Age &  (Gyr) &$8.0^{+4.1}_{-4.8}$ \\
\enddata
\end{deluxetable*}
We then estimate the stellar mass, radius, and age by modeling the spectral energy distribution (SED) using the MIST model grids \citet{Dotter2016,Choi2016} as implemented in the \texttt{EXOFASTv2} \citep{2019arXiv190709480E} package. The SED fit used Gaussian priors on the (i) 2MASS \(J,~H,~K\) magnitudes \citep{cutri_2mass_2033}, PS1 \(g^\prime,~r^\prime,~y^\prime\) PSF magnitudes from \cite{Chambers2016,Magnier2020}, and Wide-field Infrared Survey Explorer magnitudes \citep{Wright2010},\footnote{Stellar magnitudes are listed in Table~\ref{tab:SP}} (ii) spectroscopic parameters derived from \texttt{HPF-SpecMatch}, and (iii) the geometric distance calculated from \citet{Bailer-Jones2021}. The upper limit to the visual extinction is determined using estimates of Galactic dust \citep{Green2019} calculated at the distance determined by \cite{Bailer-Jones2021}. The \(R_{v}=3.1\) reddening law from \cite{Fitzpatrick1999} is employed to convert the extinction from \cite{Green2019} to a visual magnitude extinction. This fit calculates that the host star TOI-3785 has a mass of $0.52\pm{0.02}$ $M_\odot$, radius of $0.50\pm0.01$ $R_\odot$ , luminosity of $0.0367^{+0.0008}_{-0.0009}$ $L_\odot$ and an estimated age of $8.0^{+4.1}_{-4.8}$ Gyr (\autoref{tab:Parameters}). These parameters, combined with the SED derived effective temperature of $3580\pm47$ K, classifies the star as a M2-Dwarf spectral type star \citep{2016A&A...595A..95D}.

%\subsection{Stellar Activity}
To evaluate the activity of TOI-3785, we examine a Lomb-Scargle periodogram \citep{1976Ap&SS..39..447L} derived from the short cadence \textit{TESS} lightcurve. We find no significant peaks corresponding to stellar rotation. We also used the publicly available photometry from the Zwicky Transient Facility \citep[ZTF;][]{2019PASP..131a8002B} in the $g^\prime$ and $r^\prime$ filters and the All-Sky Automated Survey for Supernovae \citep[ASAS-SN;][]{2017PASP..129j4502K} in its $V$ filter. The Lomb-Scargle analysis from both sources again reports no statistically significant rotation signals in the photometry. This lack of detection is expected given our estimated $v \sin{i}$ is below our detection threshold from \texttt{HPF-SpecMatch} ($<$ 2 km/s). We further support this claim by investigating the Calcium Infrared Triplet (Ca IRT) lines \citep{1997A&AS..124..359M,2007A&A...469..309C,2005A&A...430..669A,2017A&A...605A.113M} observed by HPF and H$\alpha$ lines observed by NEID. No lines exhibited signs of emission, suggesting low activity in the chromosphere of TOI-3785 \citep{2016csss.confE...4N}. Thus, we conclude that TOI-3785 is a slowly rotating, inactive M2-dwarf star.

%\begin{figure*}[!h]
%\begin{center}
    %\textbf{SED Fit Parameter Table}\par\medskip
 % \includegraphics[scale=0.85]{tables/SED.pdf}
  %\label{SED}
 % \end{center}
  %\label{tab:SED}
%\end{figure*}

%\begin{figure*}[!h]
%\begin{center}
   % \textbf{}\par\medskip
  %\includegraphics[scale=0.9]{rvsmcmc.png}
  %\end{center}
  %\caption{TEXT HERE}
%\end{figure*}

\section{Analysis} \label{sec:analysis}

\subsection{Data Reduction}
%(Mention aperture or general ap for APO data
%backround annulus to prevent stellar contamination
%mention no detrending
%backround subtraction 
%photometric errors, taking into account inst noise and sky noise, input apo specific gain an  

We use \texttt{AstroImageJ} \citep{2016arXiv160102622C} to perform the reductions of TOI-3785's ground-based photometry. For each of the ground-based observations, we subtract a median master bias file, and for the RBO photometry, we additionally subtract a median master dark current file (at short exposure times there is no significant dark current for the ARCTIC observations). The bias and dark corrected images were then divided by their respective normalized sky flats. 

After initial data reductions were completed, we select appropriate aperture sizes for the target and reference stars to minimize both the background noise and any potential stellar interference in our photometry. We then perform differential aperture photometry using \texttt{AstroImageJ}, of the primary target and five to seven reference stars assuming a constant aperture size with a radius of 5.48\arcsec{} and 7.8\arcsec{} for APO and RBO respectively. Background values were measured by assuming a median value derived from annuli around each star with an inner and outer radius of 9.12\arcsec,~14.6\arcsec{}~for APO and 10.4\arcsec,~13.0\arcsec{}~for the RBO data. Uncertainties were calculated by \texttt{AstroImageJ} assuming photon noise from the star, background, and dark current (for RBO) and respective read noise for the individual instruments. In post-processing, we found it was unnecessary to detrend the light curves using any external parameters (airmass, background, etc.). %The reduced photometry was then packaged and used in the final \texttt{exoplanet} joint fit. 

\subsection{Joint Fitting}

Using the python package \texttt{exoplanet} \citep{exoplanet.citation}, we perform a joint fit of all transit photometry (TESS + ARCTIC + RBO) and RV measurements (HPF + NEID). We derived the final transit and radial velocity models in addition to a collection of stellar and planetary parameters that were previously estimated in the \texttt{AstroImageJ} fit.
\autoref{tab:Parameters} and \autoref{tab:exoplanet} list the finalized transit and system parameters produced by this joint fit.   

From the transit observations, we derive a best-fit a/R$_*$, impact parameter (b), transit depth (R$_p$/R$_s$)$^2$, and mid-transit ephemeris. We re-parameterized and then fit the limb darkening parameters as suggested in  \citet{kipping.limbdark} to ensure uninformative sampling of quadratic parameters. As each instrument employs a different bandpass, we fit for individual quadratic limb darkening terms. We also include a photometric noise jitter term added in quadrature to the error bars and the addition of a flux offset value to each light curve. 

Due to \textit{TESS}'s large pixel sizes, photometric dilution is a common source of error in transit depth estimations \citep{2015ApJ...809...77S}. \textit{TESS} dilution may cause photometric variation in our reported transit depth causing our errors to inflate. We account for this by fitting a separate dilution term multiplied to the transit depth for each \textit{TESS} Sector \citep[as described in][]{2020MNRAS.499.3139B}). For \textit{TESS} Sector 20, we measure a dilution factor of $D_{TESS_{20}} = 0.97^{+0.06}_{-0.05}$ and for \textit{TESS} Sector 47 a dilution factor of $D_{TESS_{47}} = 0.92\pm 0.05$. Our high-precision uncontaminated ground-based photometry from ARCTIC (which we use as the baseline fixed to a dilution of 1) enabled us to properly account for this variation.

\autoref{fig:LC} displays our best-fit photometric transit models. These folded light curves report a transit depth of 0.9254$\pm0.0003 $\% and transit duration of $T_{duration}$ = 0.071$\pm 0.001$ days ($\sim$1.7 hours). Each transit plot presents a 10-minute bin of the reduced data and residuals as well as values of median photometric error.  

For the radial velocity observations, we include linear RV trend terms for both HPF and NEID to account for any slight positive or negative slopes in the RVs caused by instrumental drift. In addition, we report the instrument-specific factors of RV jitter and offset. The jitter term is used to estimate the degree of RV error inflation in order to meet an expected RV fit. All photometric and radial velocity correction terms are reported in \autoref{tab:exoplanet}. We plot, in \autoref{fig:RVcurve}, the \texttt{exoplanet} RV fit including all HPF and NEID points. The best-fit model indicates an RV semi-amplitude of 9.24 $\pm$ 2.68 m/s and an eccentricity of $e = 0.11^{+0.10}_{-0.08}$.  From this analysis, we determine that TOI-3785~b has a radius of $5.14 \pm 0.16~R_\oplus$ and a mass of $14.95^{+4.10}_{-3.92}~M_\oplus$. \autoref{tab:Parameters} lists the finalized planetary and orbital parameters produced by this joint fit.

\begin{deluxetable*}{lrrrrr}[htbp]
\centering
\caption{Photometry/Radial Velocity Correctional Terms}
\label{tab:exoplanet}
\tabletypesize{\scriptsize}
\tablehead{\colhead{Parameter} & \colhead{Label (Units)} & \colhead{Value} & & &}
\startdata
%divide by instrument instead of variable
%are limb darkening coeff re-parameterized?
\textbf{Photometric Parameters:} & &\textbf{TESS S20} &\textbf{TESS S47} &\textbf{APO} &\textbf{RBO} \\
\quad Linear Limb-Darkening Coefficient &$u_1$ &0.36$^{+0.39}_{-0.26}$ &0.26$^{+0.31}_{-0.19}$ &0.27$^{+0.26}_{-0.19}$ &0.35$^{+0.43}_{-0.25}$ \\
\quad Quadratic Limb-Darkening Coefficient &$u_2$ &0.17$^{+0.37}_{-0.37}$ &0.14$^{+0.32}_{-0.27}$ &0.046$^{+0.268}_{-0.214}$ &0.10$^{+0.37}_{-0.31}$ \\
\quad Photometric Jitter &$\sigma_{phot}$(ppm) &53$^{+55}_{-33}$ &110$^{+103}_{-67}$ &3056$^{+98}_{-91}$ &3830$^{+430}_{-360}$ \\
\quad Dilution Factor &D &0.97$^{+0.06}_{-0.05}$ &0.92$\pm 0.05$ &*** &*** \\
\hline %\midrule
\textbf{RV Parameters:} & &\textbf{HPF} &\textbf{NEID} & & \\
\quad RV Jitter &$\sigma_{RV}$ (m/s) &$2.9^{+3.1}_{-2.1}$ &$7.7^{+3.5}_{-2.6}$ & & \\
\quad RV Offset &$\gamma_{RV}$ (m/s) &4.7$\pm{3.3}$ &1.1$\pm{2.9}$ & & \\
\hline
\quad RV Trend\tablenotemark{a} & $\dot{\gamma}$ (mm/s/day) &-2.1$^{+4.2}_{-4.3}$ & & & \\
\quad Absolute RV\tablenotemark{a} & $\Delta RV$ (m/s) &4657$\pm$152 & & & \\
\enddata
\tablenotetext{a}{Not instrument specific}
\end{deluxetable*}

\subsection{Planetary Companions}
Further analyzing the transit data, we search for additional periodic signals by implementing a Box Least Squared \citep[BLS;][]{BLS} algorithm of all available \textit{TESS} data extracted from the MAST archive \citep{MAST}. The known transit of TOI-3785~b in both Sectors 20 and 47 were masked to twice the duration in order to search for other potential period detections that may indicate additional transiting planets in the TOI-3785 system. The masked BLS periodograms report no significant peaks over the false alarm probability (FAP) of 10\%.     

Additionally, periodograms of HPF and NEID also show no additional significant peaks (stellar or planetary in nature). The existing data does not reveal the presence of a close-period companion in this system as the known signal of TOI-3785~b recovers the lowest FAP at ~10\%. However, the limited coverage of this system with both \textit{TESS} and RV monitoring cannot rule out the potential for additional long-period planets. From our available transit and RV data, we see no detection of additional orbiters, but a more in-depth analysis is required for a concrete claim to be made.

\section{Discussion} \label{sec:discussion}
\subsection{TOI-3785~b in M-dwarf Parameter Space}
\begin{figure*}[!h]
    \begin{minipage}[b]{\linewidth}
    \centering
    \hspace*{0cm}
    \includegraphics[scale=0.7]{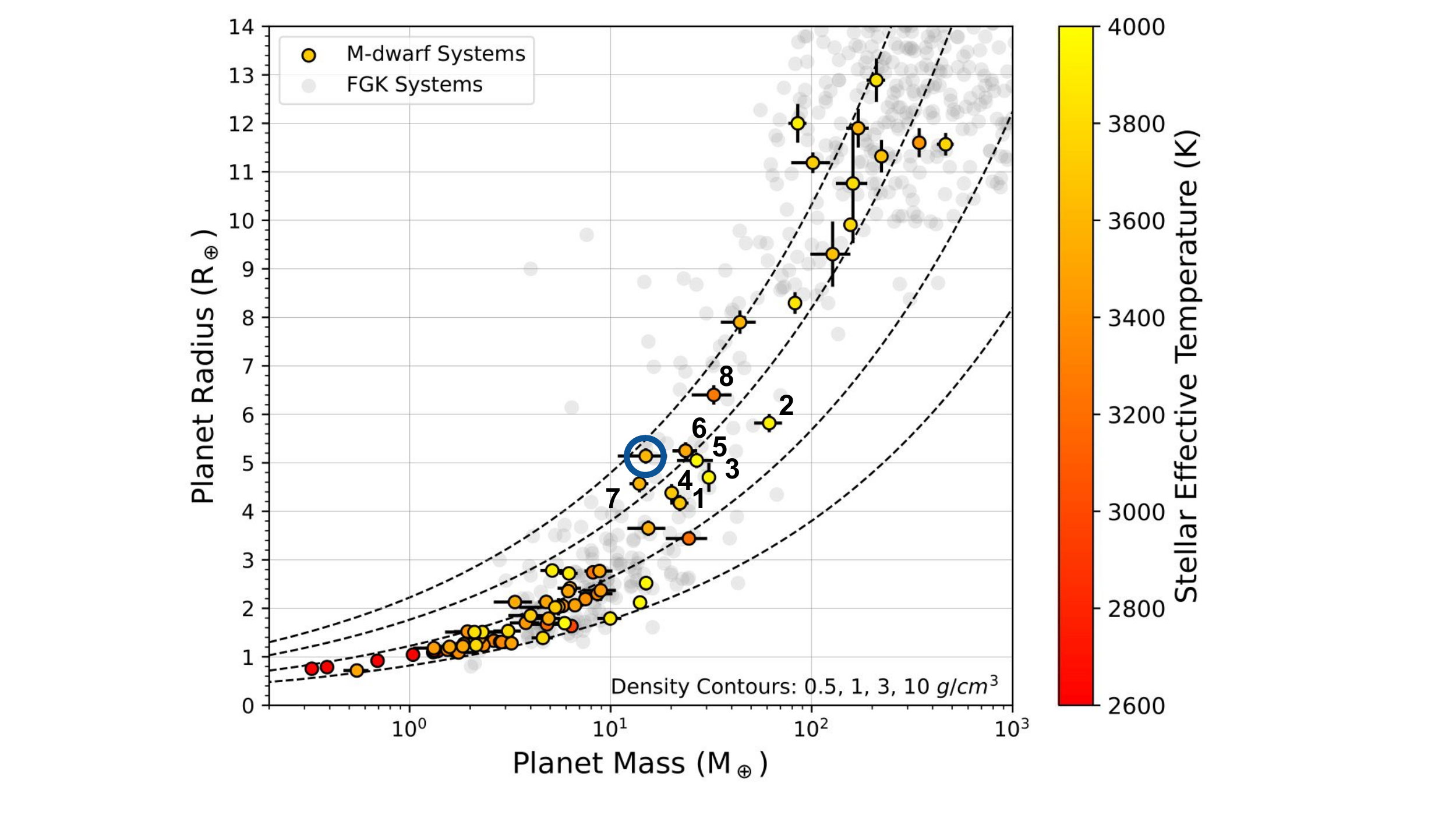}
    \end{minipage}
    \begin{minipage}[b]{\linewidth}
    \hspace*{-1cm}
    \centering
    \includegraphics[scale=0.66]{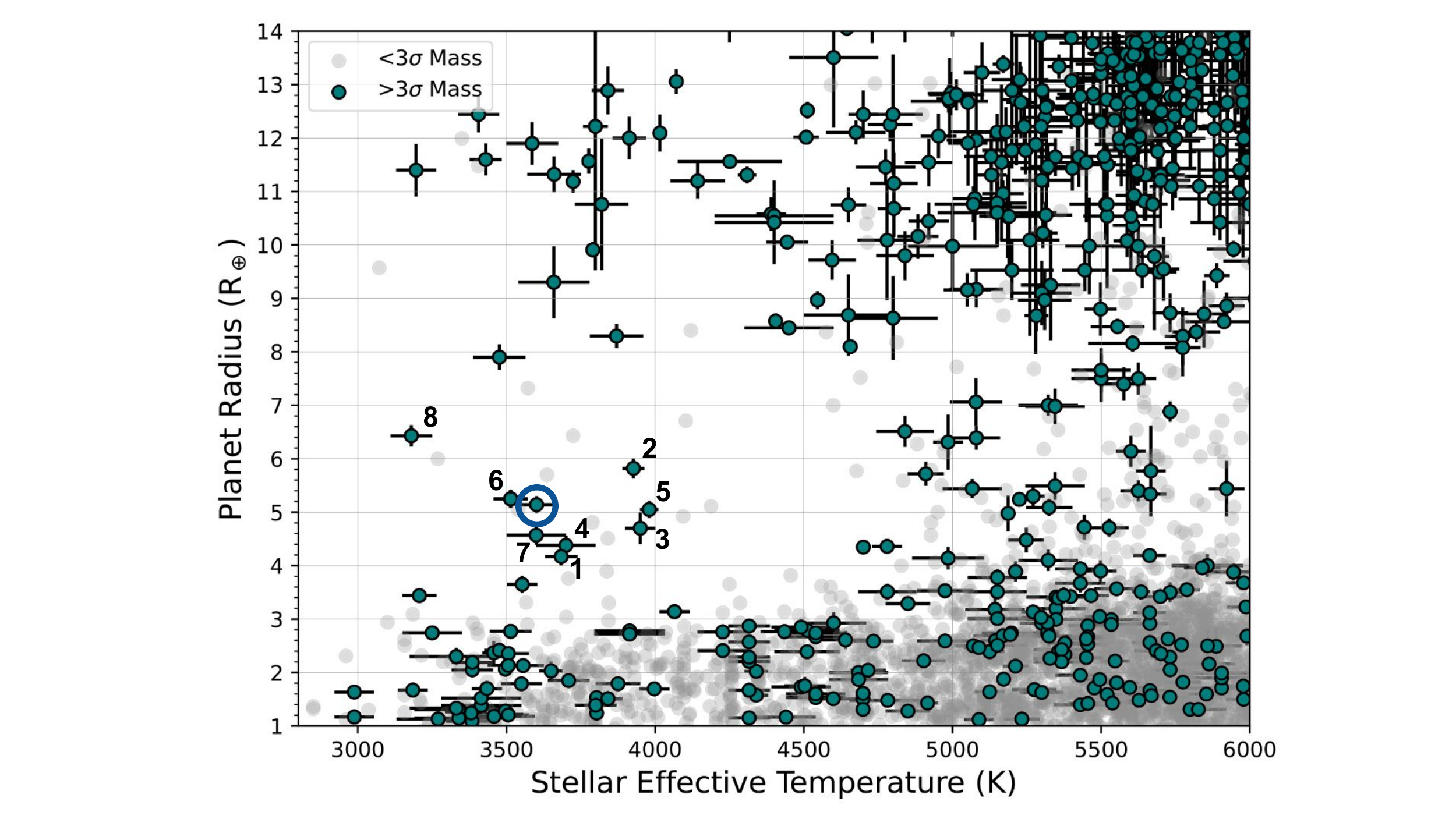}
    \end{minipage}
    \caption{\textit{Top}: Planet radius against planet mass of only M-dwarf targets (restricted by an upper bound on effective temperature  of $T_{\rm{eff}} < 4000 K$). The points with attributed effective temperatures are M-dwarf planets that have $>3\sigma$ mass precision. Gray points represent $>3\sigma$ FGK targets. The dashed lines are density contours with specific values listed in the bottom right. \textit{Bottom}: Planetary radius against stellar effective temperature, to place TOI-3785~b against a broader range of stellar host types. Colored points represent $>3\sigma$ mass precision while targets under this threshold are grayed out. We also highlight the lack of planetary confirmation between (4000 K $<$ $T_\mathrm{eff}$ $<$ 4700 K). This is most likely a result of the shallow transit depths that are characteristic of K-dwarf stars. This creates a detection bias as K-dwarf planets produce weaker transiting signals. In both figures, similar M-dwarf Neptunes to TOI-3785~b are labeled and listed in \autoref{tab:SIMILAR}.}
    \label{fig:MR}
\end{figure*}

In order to emphasize the unique planetary characteristics of TOI-3785~b we compare this system to other confirmed exoplanet targets in \autoref{fig:MR}. We plot TOI-3785~b in both planetary Mass-Radius (\autoref{fig:MR}; top) and $T_{eff}$-Radius (\autoref{fig:MR}; bottom) space. These systems were compiled from the NASA Exoplanet Archive \citep{Archive} as of 2023 March 1 using the following parameter constraints: an upper planetary radius limit of $14$ $R_\Earth$ and a radius and mass significance cut off at $>$ 3$\sigma$. For the Mass-Radius plot we limit the stellar effective temperature to $<$ 4000 K \citep[the upper temperature boundary of M-dwarfs][]{casagrande.mdwarfs} and include planetary density contour lines at 0.5, 1, 3, and 10 $g/cm^3$.

%Additionally, \citet{gan.mdwarf.occurence} derives an occurrence rate of M-dwarf targets of higher radii such as 0.3\% for M-dwarf Hot-Jupiters - slightly smaller than the $\sim$1\% predicted for FGK stars \citep[e.g.,][]{2020MNRAS.492..377W}.
The parameter spaces of \autoref{fig:MR} both show TOI-3785~b, indicated by a blue circle, to be a meaningful addition to the current number of known M-dwarf-hosted targets. Due to its planetary radius (4 $R_\Earth$ $<$ $R_p$ $<$ 7 $R_\Earth$), TOI-3785~b occupies the rare M-dwarf Neptune population of which only 8 others have been confirmed with a $> 3\sigma$ mass and radius (\autoref{fig:MR}). This dearth in the Neptune population becomes clear when the vast number of lower radii targets ($< 3R_\Earth$) is considered. It is widely known that lower mass M-dwarfs have a higher occurrence rate for smaller (and likely terrestrial in composition) close-in planets with \citet{dressing.mdwarf.occurrence} discovering a sharp decrease in occurrence rate at 4 $R_\Earth$. 

The M-dwarf Jupiter population ($>$ 7 $R_\Earth$) is seen to be relatively sparse compared to FGK occurrence totaling only $\sim$15 mass significant targets. M-dwarf Jupiters do not come close to rivaling the M-dwarf Earth population ($<$ 3 $R_\Earth$) in which $\sim$35 $>3\sigma$ mass targets are known. Still, first approximations of 
occurrence rates have been derived for close-in Jupiters orbiting M-dwarfs, even with this small sample size \citep[$\sim$1\%;][]{gan.mdwarf.occurence,2023MNRAS.521.3663B}. However, the occurrence of M-dwarf Neptunes (4 $<$ $R_p$ $<$ 7 R$_{\oplus}$) has yet to be the focus of a targeted study - in part due to the $<$ 10 confirmed detections. TESS’s focus on nearby M-dwarfs is steadily growing the Neptune population. With the discovery of additional Neptunes similar to TOI-3785~b, we may soon derive the first occurrence rates for Neptunes and move closer towards a complete picture regarding the occurrence of all M-dwarf populations.
%Currently, TOI-3785~b is only one of ~10 confirmed Neptunes in this radius range around M-dwarfs with a measured ($>3\sigma$) mass and radius (\autoref{fig:MR}; top). However, due to TESS’s focus on nearby M-dwarfs, this population is steadily growing. With the discovery of additional Neptunes similar to TOI-3785~b, we can soon bridge the current investigations into occurrence rates of Jupiters and Earths with those of Neptunes around M-dwarfs. .As previously mentioned, only upper limit constraints have been placed on M-dwarf Jupiters while no published investigations into Neptune occurrence rates have been conducted.       

\subsection{Constraints on M-dwarf Planetary Formation}
The leading theory of Neptune formation around M-dwarfs is core accretion \citep{laughlin.2004}, in which the formation of a solid core generates a disk of gas and dust from surrounding debris that is slowly accreted onto the surface of the protoplanet. In cases of ample material and undisturbed mass accumulation, a protoplanet may reach a critical mass triggering runaway accretion in which a planet's mass exponentially increases. This is the traditional formation pathway for many Jupiter mass planets \citep[e.g.,][]{bodenheimer.coreaccretion,Pollack1996}. In the case of the less massive Neptune population, there must be an inhibitor to prevent runaway accretion from taking place: either the protoplanet lacks sufficient material to accrete onto its surface or it lacks sufficient time to grow to a critical mass \citep{NEPformation}. \citet{laughlin.2004} argues that due to the smaller M-dwarf disk masses, gas giant cores require additional time to form. If they reach the critical mass threshold to begin accumulating H/He, their runaway growth stage is cut short due to disk dispersion.

TOI-3785~b appears to support this formation theory. Using the Exoplanet Compositional Interpolator\footnote{https://tools.emac.gsfc.nasa.gov/ECI/} based on models from \citet{lopez.and.fortney.2014}, we estimate a H/He mass fraction of 20\% ($\sim$3 M$_\oplus$) with a heavy-element (core) mass fraction of 80\% (11.95 M$_\oplus$) for TOI-3785~b. As this is slightly more massive than the predicted core mass required for runaway accretion, we conclude that TOI-3785~b’s core must have formed slowly following the predicted pathway highlighted in \citet{laughlin.2004}. With 20\% of its mass in a H/He envelope, it appears TOI-3785~b was poised to begin runaway accretion. However, this accretion stalled potentially due to the disk dispersing or the planet migrating inwards to its present-day location. By further investigating TOI-3785~b’s composition, we may constrain the formation timeline of M-dwarf hosting Neptunes to derive reasonable evolutionary pathways for these rare targets.

\subsection{The Neptune Desert}
\begin{figure*}[!h]
    \begin{minipage}[b]{\linewidth}
    \centering
    \includegraphics[scale=0.65]{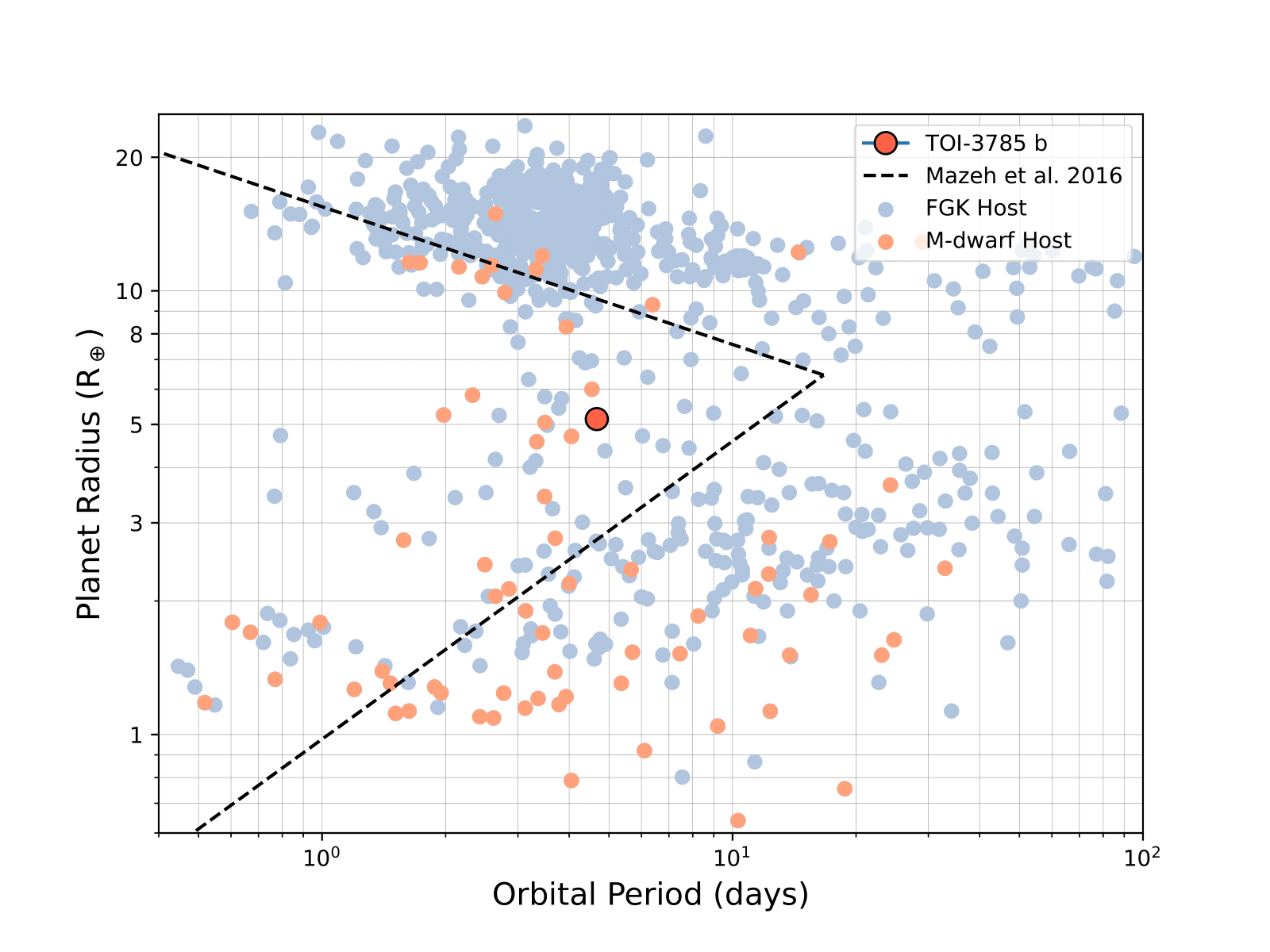}
    \end{minipage}
    \begin{minipage}[b]{\linewidth}
    \hspace*{-0.1cm}
    \centering
    \includegraphics[scale=0.70]{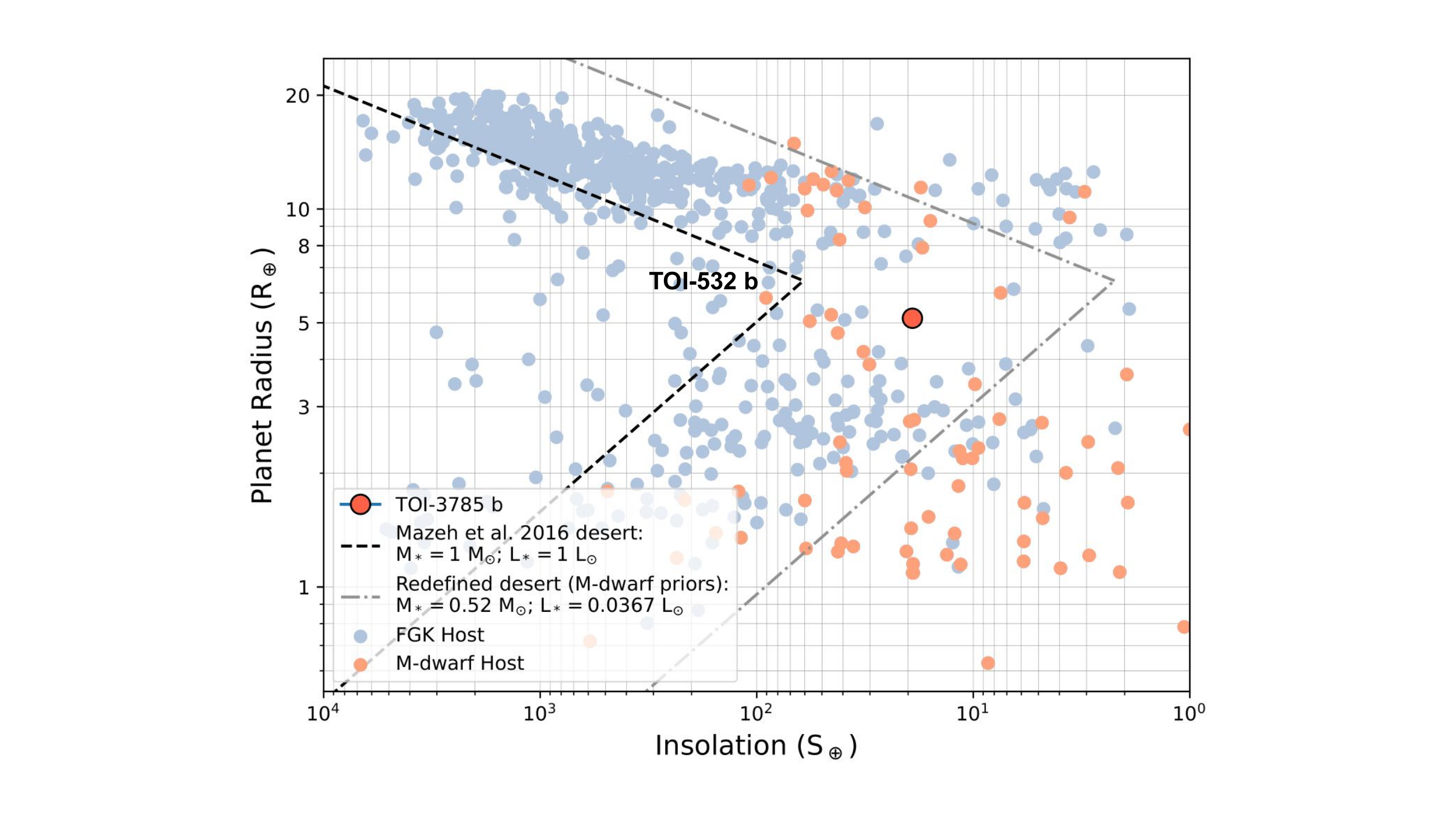}
    \end{minipage}
    \caption{\textit{Top}: The Neptune Desert as reported by \citep{DerthNep} in orbital period  - planetary radius space. Indicated by the red dot is TOI-3785~b. \textit{Bottom}: We reorient the Neptune Desert bounds to apply to Insolation - Radius space as per host star characteristics. The dark lines represent the desert for sun-like stars which define appropriate bounds for FGK hosts. The gray lines show a redefined desert that considers the lower mass and luminosity of M-dwarf hosts, so much so that this boundary is no longer defined as a notable desert. If we instead use the insolation-based definition (black dashed line), only one planet around an M-dwarf, TOI-532 b (see \autoref{tab:SIMILAR} [2]), falls inside the Neptune Desert. In both cases, only $>3\sigma$ mass targets are considered.}    
    \label{fig:DES}
\end{figure*}

 The Neptune desert is a region of parameter space in which remarkably few Neptune-sized targets have been confirmed around FGK stars. Targets that inhabit this region are defined by their Neptune radii as well as low orbital periods and high insolations. TOI-3785~b lies within the Neptune-desert regime as defined in Radius-Period space in~\cite{DerthNep} (\autoref{fig:DES}: \textit{top}). However, the bounds of this desert were derived from FGK targets confirmed by the Kepler mission. M-dwarf targets, such as TOI-3785~b, may have misleading placements within the desert as low-temperature stars will produce planets with low insolation even at short periods. TOI-3785~b possesses a short orbital period of 4.67 days and its cooler host star yields a significantly smaller insolation (19x Earth Insolation) when compared to planets around FGK stars that normally yield insolations within the range of 100 - 1000 $S_\Earth$. Therefore, the Neptune Desert should be considered in Radius-Insolation space for M-dwarf hosting systems. TOI-3785~b sits outside of the Insolation Neptune Desert space as defined by \citet{Kanodia532}(\autoref{fig:DES}:\textit{bottom}). 
 
TOI-532~b \citep{Kanodia532} is a similarly sized planet compared to TOI-3785~b orbiting a slightly larger M0-dwarf. TOI-532~b is also the only M-dwarf Neptune that possesses a large enough insolation (94 S$_\oplus$) to be considered within the FGK bounds of the insolation Neptune Desert. While similarly sized, this planet possesses a substantially higher mass and density than all other M-dwarf Neptunes suggesting it experienced significant H/He escape during its lifetime. With its low insolation, TOI-3785~b likely experienced little to no atmospheric escape during its evolution. %Therefore, even though this planet falls in the Neptune Desert defined in period space, we suggest the need for caution when comparing the atmospheric evolution of Neptunes around FGK stars and those around M-dwarfs as the environments they evolved in were likely drastically different at comparable periods.

\subsection{Atmospheric Comparative Planetology}

\begin{figure*}[ht]
    \centering
    \hspace*{-1cm}
    \includegraphics[scale=0.8]{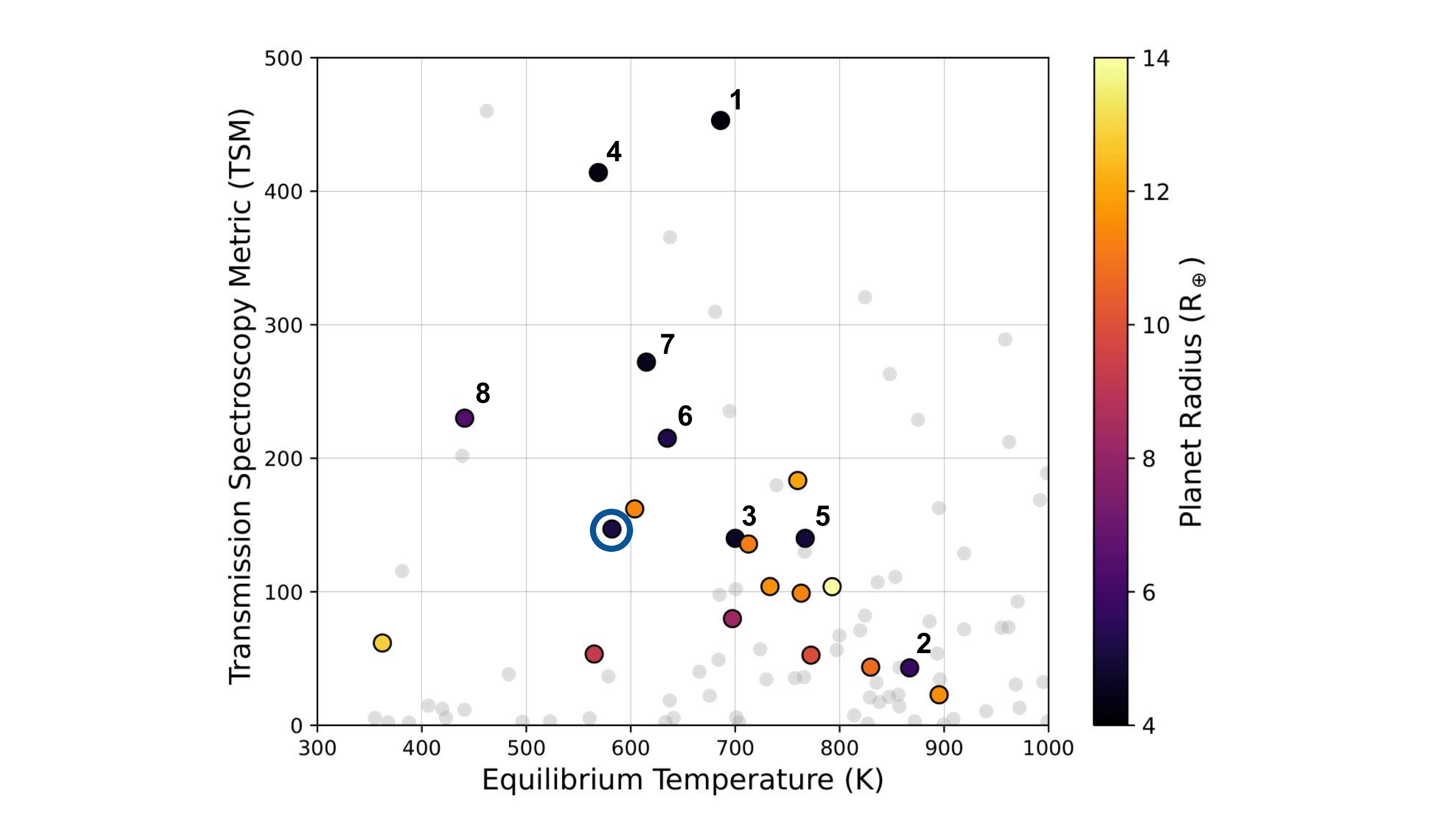}
    \caption{\textbf{The transmission spectroscopy metric} (TSM) of known exoplanets with both a measured ($>$ 3$\sigma$) mass and radii $>4~R_\oplus$ as a function of equilibrium temperature. Planets orbiting M-dwarfs are labeled according to \autoref{tab:SIMILAR} and are noted with colors varying by their respective planetary radii, while planets orbiting FGK dwarf stars are plotted as gray points. Notable M-dwarf planets are labeled. TOI-3785~b (navy blue circle) possesses one of the highest TSMs for any planet cooler than 600K, making it a promising target for future transmission spectroscopy observations.}
    \label{fig:TSM}
\end{figure*}

TOI-3785~b joins the growing list of promising targets for atmospheric characterization (\autoref{tab:SIMILAR}). We calculate its Transmission Spectroscopy Metric (TSM) following Equation (1) in \citet{kempton.tsm} finding a TSM value of 147 (\autoref{fig:TSM}). TOI-3785~b possesses one of the highest TSMs for any planet cooler than 600 K--with AU Mic b the only other planet in this temperature regime with a higher TSM. Since AU Mic is an active star \citep{plavchan.aumicb}, difficulties probing AU Mic b's atmosphere may arise due to stellar interference. Thus, the inactivity of its host star makes TOI-3785~b the best target for exploring this temperature regime.  

TOI-3785~b also possesses similar planetary and stellar parameters as two Neptunes with well-characterized transmission spectra: GJ 3470~b \citep[e.g.,][]{crossfield.gj3470.2013,ehrenreich.gj3470.2014,dragomir.gj3470.2015} and GJ 436~b \citep[e.g.,][]{knutson.gj436b.2014}. Both of these targets possess featureless spectra within Hubble's Wide Field Camera 3 bandpass (1.1--1.7 microns) indicating hazy atmospheres; a characteristic that TOI-3785~b could share \citet{yu.haze.trends,dymont.hazes}. By leveraging JWST's NIRSpec-Prism longer wavelength coverage, the hazes should become translucent at wavelengths beyond 3 microns allowing for both characterization of the haze layer and the atmospheric composition beneath \citep{kawashima.hazes}. TOI-3785~b therefore presents an opportunity to not only explore the atmosphere of a warm-Neptune but also enable insightful atmospheric comparisons with similar planets around similar stars. Interestingly, both GJ 3470~b and GJ 436~b have escaping atmospheres observed via helium \citep[for GJ 3470~b][]{ninan.gj3470.helium} or Lyman-alpha \citep[for GJ 436~b][]{ehrenreich.lymanalpha.gj436} absorption features. TOI-3785~b's low-density of $\sim$0.6 $g/cm^{3}$ along with similarities between this system and GJ 3470 (stellar parameters) makes it a promising target for helium follow-up. %We observed one successive spectra derived from in-transit HPF points whose wavelength range covers the helium 10830$\angstrom$ triplet. We observe no significant helium absorption feature during transit when compared to the out-of-transit spectra. With similar stellar hosts, it is unclear why TOI-3785~b does not appear to possess an escaping atmosphere. Future targeted helium observations confirming this null detection are required, but this apparent deviation from similar planets in this population suggests a new window into atmospheric escape.

\begin{deluxetable*}{lrrrrrrr}[htbp]
\centering
\caption{Similar M-dwarf Neptune Targets Ordered by Planet Density}
\label{tab:SIMILAR}
\tabletypesize{\footnotesize}
\tablehead{\colhead{Planet / Source} & \colhead{$T_{\rm{eff}}$ (K)} & \colhead{TSM} & \colhead{$T_{EQ}$ (K)} & \colhead{$S_p$ ($S_\Earth$)} & \colhead{$M_p$ ($M_\oplus$)} & \colhead{$R_p$ ($R_\oplus$)} &$\rho_p$ ($g/cm^3$)}
\startdata 
\textbf{[1]}  GJ 436 b\\\citet{2014AcA....64..323M} &3586$\pm$36 &453&686$\pm$10&41$\pm$4 &22.1$\pm 2.3$ &4.170$\pm$0.168& 1.80$\pm$0.29\\
\hline
\textbf{[2]}  TOI-532 b\\\citet{Kanodia532}  &3927$\pm$37 &43&867$\pm$18& 94.10$\pm$8.00&61.5$^{+9.7}_{-9.3}$ &5.82$\pm$0.19 &1.72 $\pm$ 0.31\\
\hline
\textbf{[3]}  LP 714-47 b \\\citet{LP47b}&3950$\pm$51 &140&$700^{+19}_{-24}$& 46$\pm$2 &30.8$\pm$1.5 &4.7$\pm$0.3& 1.7$\pm$0.3 \\
\hline
\textbf{[4]}  AU Mic b*\\\citet{plavchan.aumicb} &3700$\pm$100 &414&569.5$_{-4.5}^{+5.1}$&22$\pm$2 &20.12$^{+1.57}_{-1.72}$ &4.38$\pm$0.18 &$1.32^{+0.19}_{-0.20}$ \\
\hline
\textbf{[5]}  TOI-1728 b \\\citet{Kanodia1728}&3980$\pm$32 &130&767$\pm$8&57.78$\pm$3.48 &26.78$^{+5.43}_{-5.13}$ &5.05$^{+0.16}_{-0.17}$ & $1.14^{+0.26}_{-0.24}$ \\
%K2-14b &3789 $\pm$ 17 &8.36 &*** &4.18$\pm$0.42 &\citep{K2-14b} \\
%K2-43b &3840 $\pm$ $_{-50}^{+98}$ &3.47 &*** &4.51 $_{-0.19}^{+0.44}$ &\citep{K2-43b} \\
\hline
\textbf{[6]}  TOI-674 b \\\citet{Murgas674}&3514$\pm$57 &215&635$\pm$15&38.4$\pm$0.1 &23.6$\pm$3.3 &5.25$\pm$0.17 &0.91$\pm$0.15\\
\hline
\textbf{[7]}  GJ 3470 b \\\citet{Awiphan} &3622$_{-55}^{+58}$ &272 &615$\pm$16&42$\pm$6&13.9$\pm$1.5 &4.57$\pm$0.18 &0.80$\pm$0.13 \\
\hline
\textbf{[8]}  TOI-3884 b*\\\citet{almenara.toi3884}\\\citet{3884} &3180$\pm 88$ &230 &441$\pm$15&6.29$\pm$0.84&32.59$^{+7.31}_{-7.38}$ &6.43$\pm$0.20 &0.67$^{+0.18}_{-0.16}$
\\
\hline
\textbf{TOI-3785~b} \\\textbf{This Work}&\textbf{3576}$\mathbf{\pm 88}$ &\textbf{147}&\textbf{582}$\mathbf{\pm 16}$& \textbf{19.1}$\mathbf{\pm 2.0}$&\textbf{14.95}$\mathbf{^{+4.10}_{-3.92}}$ &\textbf{5.14} $\mathbf{\pm 0.16}$ &\textbf{0.61}$\mathbf{^{+0.18}_{-0.17}}$\\ \hline
\enddata
\tablenotetext{*}{Active Host Star}
\end{deluxetable*}

\section{Summary} \label{sec:summary}
Using both ground-based and \textit{TESS} transit photometry as well as spectroscopic RV follow-up of the TOI-3785 system, we confirm the existence of a single planetary companion, TOI-3785~b, a warm-Neptune with a 4.67 day circular orbit around an M2V-dwarf star. Using the package \texttt{exoplanet} we model both the transit observations and RVs to derive a planetary mass of 14.95$^{+4.10}_{-3.92} M_\oplus$ and radius of $5.14 \pm 0.16 R_\oplus$. The confirmation of TOI-3785~b proves to be a valuable addition to the small number of M-dwarf-hosted Neptunes as increased target confirmation in this space may support Neptune formation models such as the joint efforts of core accretion and situational disk dispersion. Future investigations into this target via transmission spectroscopy are warranted as it possesses an ideal TSM along with favorable constraints on atmospheric hazing. The noteworthy similarities to GJ 3470~b and GJ 436~b may also demonstrate similarly influential results on the composition and formation pathways of M-dwarf gas planets. Furthermore, we discuss TOI-3785~b's place relative to the radius-period and radius-insolation Neptune Deserts and the necessary cautions that accompany M-type hosts and desert classifications.   %A non-detection of a He 10830$\angstrom$ exosphere supports our claim that TOI-3785~b is observed outside of the radius-isolation Neptune Desert.                          

%From in transit RV points we claim a non-detection of He 10830$\angstrom$ which points towards an inactive host star, which is consistent with the inactivity seen in the Lomb-Scargle analysis. Given the inactive nature of TOI-3785, in addition to high TSM and relatively low atmospheric hazing, we propose transmission spectroscopy via JWST as an appropriate means to further this target's analysis. Probing the atmosphere of a target such as this may reveal compositional characteristics unique to that of low-density warm-Neptunes. Thus revealing potential indicators as to how this target formed and evolved into its current state. By confirming additional M-dwarf Neptunes we move closer to quantifying the occurrence rate of these rare planets as well as further the investigation of the formation and evolutionary tendencies of Neptune-like planets.          

\section{Acknowledgments}
%We thank the anonymous referee for valuable feedback which has
%improved the quality of this manuscript.\\
% \\
We want to thank the anonymous referee for their thoughtful suggestions and time spend on this work. Their comments regarding the specifics of planetary occurrence and Neptune detection were helpful in shaping the discussion of this planetary target.  
% CEHW / ACI / PSU
The Center for Exoplanets and Habitable Worlds is supported by Penn State and the Eberly College of Science.
Computations for this research were performed on the Penn State’s Institute for Computational and Data Sciences' Advanced Cyber Infrastructure (ICDS-ACI).  This content is solely the responsibility of the authors and does not necessarily represent the views of the Institute for Computational and Data Sciences.
The Pennsylvania State University campuses are located on the original homelands of the Erie, Haudenosaunee (Seneca, Cayuga, Onondaga, Oneida, Mohawk, and Tuscarora), Lenape (Delaware Nation, Delaware Tribe, Stockbridge-Munsee), Shawnee (Absentee, Eastern, and Oklahoma), Susquehannock, and Wahzhazhe (Osage) Nations.  As a land grant institution, we acknowledge and honor the traditional caretakers of these lands and strive to understand and model their responsible stewardship. We also acknowledge the longer history of these lands and our place in that history.

WDC acknowledges support from NSF grant AST 2108801.

% HPF / HET
These results are based on observations obtained with the Habitable-zone Planet Finder Spectrograph on the HET. We acknowledge support from NSF grants AST-1006676, AST-1126413, AST-1310885, AST-1310875, AST-1910954, AST-1907622, AST-1909506, ATI 2009889, ATI-2009982, AST-2108512, and the NASA Astrobiology Institute (NNA09DA76A) in the pursuit of precision radial velocities in the NIR. The HPF team also acknowledges support from the Heising-Simons Foundation via grant 2017-0494.  The Hobby-Eberly Telescope is a joint project of the University of Texas at Austin, the Pennsylvania State University, Ludwig-Maximilians-Universität München, and Georg-August Universität Gottingen. The HET is named in honor of its principal benefactors, William P. Hobby and Robert E. Eberly. The HET collaboration acknowledges the support and resources from the Texas Advanced Computing Center. We thank the Resident Astronomers and Telescope Operators at the HET for the skillful execution of our observations with HPF. We would like to acknowledge that the HET is built on Indigenous land. Moreover, we would like to acknowledge and pay our respects to the Carrizo \& Comecrudo, Coahuiltecan, Caddo, Tonkawa, Comanche, Lipan Apache, Alabama-Coushatta, Kickapoo, Tigua Pueblo, and all the American Indian and Indigenous Peoples and communities who have been or have become a part of these lands and territories in Texas, here on Turtle Island.

% KP / WIYN / NEID / NESSI
Based on observations at Kitt Peak National Observatory, NSF’s NOIRLab, managed by the Association of Universities for Research in Astronomy (AURA) under a cooperative agreement with the National Science Foundation. The authors are honored to be permitted to conduct astronomical research on Iolkam Du’ag (Kitt Peak), a mountain with particular significance to the Tohono O’odham. Deepest gratitude to Zade Arnold, Joe Davis, Michelle Edwards, John Ehret, Tina Juan, Brian Pisarek, Aaron Rowe, Fred Wortman, the Eastern Area Incident Management Team, and all of the firefighters and air support crew who fought the recent Contreras fire. Against great odds, you saved Kitt Peak National Observatory.

Data presented herein were obtained at the WIYN Observatory from telescope time allocated to NN-EXPLORE through the scientific partnership of the National Aeronautics and Space Administration, the National Science Foundation, and the National Optical Astronomy Observatory. WIYN is a joint facility of the University of Wisconsin–Madison, Indiana University, NSF’s NOIRLab, the Pennsylvania State University, Purdue University, University of California, Irvine, and the University of Missouri.

Data presented were obtained by the NEID spectrograph built by Penn State University and operated at the WIYN Observatory by NSF's NOIRLab, under the NN-EXPLORE partnership of the National Aeronautics and Space Administration and the National Science Foundation.  This work was performed for the Jet Propulsion Laboratory, California Institute of Technology, sponsored by the United States Government under the Prime Contract 80NM0018D0004 between Caltech and NASA. These results are based on observations obtained with NEID under proposals 2021B-0035 (PI: S. Kanodia), 2021B-0435 (PI: S. Kanodia), 2022A-452266 (PI: S. Kanodia), and 2022A-794607 (PI: J. Libby-Roberts). We thank the NEID Queue Observers and WIYN Observing Associates for their skillful execution of our NEID observations.

Some of the observations in this paper made use of the NN-EXPLORE Exoplanet and Stellar Speckle Imager (NESSI). NESSI was funded by the NASA Exoplanet Exploration Program and the NASA Ames Research Center. NESSI was built at the Ames Research Center by Steve B. Howell, Nic Scott, Elliott P. Horch, and Emmett Quigley.

The WIYN Observatory is a joint facility of the NSF's National Optical-Infrared Astronomy Research Laboratory, Indiana University, the University of Wisconsin-Madison, Pennsylvania State University, the University of Missouri, the University of California-Irvine, and Purdue University.
% APO / ARCTIC / diffuser
The ground-based photomety is based on observations obtained with the Apache Point Observatory 3.5-meter telescope, which is owned and operated by the Astrophysical Research Consortium. We acknowledge support from NSF grant AST-1907622 in the pursuit of precise photometric observations from the ground.

% ZTF
Based on observations obtained with the Samuel Oschin 48-inch Telescope at the Palomar Observatory as part of the Zwicky Transient Facility project. ZTF is supported by the National Science Foundation under Grant No. AST-1440341 and a collaboration including Caltech, IPAC, the Weizmann Institute for Science, the Oskar Klein Center at Stockholm University, the University of Maryland, the University of Washington, Deutsches Elektronen-Synchrotron and Humboldt University, Los Alamos National Laboratories, the TANGO Consortium of Taiwan, the University of Wisconsin at Milwaukee, and Lawrence Berkeley National Laboratories. Operations are conducted by COO, IPAC, and UW.

% SIMBAD /ADS
This research has made use of the SIMBAD database, operated at CDS, Strasbourg, France, and NASA's Astrophysics Data System Bibliographic Services. 
% EXOFOP
This research has made use of the Exoplanet Follow-up Observation Program website, which is operated by the
1197 California Institute of Technology, under contract with the National Aeronautics and Space Administration under the Exoplanet Exploration Program.
% NASA Exoplanet Archive
This research has made use of the NASA Exoplanet Archive, which is operated by Caltech, under contract with NASA under the Exoplanet Exploration Program.

Some of the data presented in this manuscript were obtained from the Mikulski Archive for Space Telescopes (MAST) at the Space Telescope Science Institute. The specific observations analyzed can be accessed via \citet{MAST}

% TESS / MAST
Some of the data presented in this paper were obtained from MAST at STScI. Support for MAST for non-HST data is provided by the NASA Office of Space Science via grant NNX09AF08G and by other grants and contracts. This work includes data collected by the TESS mission, which are publicly available from MAST. Funding for the TESS mission is provided by the NASA Science Mission directorate.

% Gaia
This work presents results from the European Space Agency (ESA) space mission Gaia. Gaia data are being processed by the Gaia Data Processing and Analysis Consortium (DPAC). Funding for the DPAC is provided by national institutions, in particular the institutions participating in the Gaia MultiLateral Agreement (MLA). The Gaia mission website is https://www.cosmos.esa.int/gaia. The Gaia archive website is https://archives.esac.esa.int/gaia.

%Exofop
This research has made use of the Exoplanet Follow-up Observation Program (ExoFOP; DOI: 10.26134/ExoFOP5) website, which is operated by the California Institute of Technology, under contract with the National Aeronautics and Space Administration under the Exoplanet Exploration Program.

CIC acknowledges support by NASA Headquarters through an appointment to the NASA Postdoctoral Program at the Goddard Space Flight Center, administered by USRA through a contract with NASA and the NASA Earth and Space Science Fellowship Program through grant 80NSSC18K1114.

\facilities{TESS, APO (ARCTIC), RBO, Gaia, HET (HPF), WIYN (NEID, NESSI)}

\newpage
\bibliographystyle{aasjournal}
\bibliography{TOI-3785paper}


\end{document}